\documentclass[twocolumn]{aastex61}

\usepackage[utf8]{inputenc}
\usepackage{hyperref}
\usepackage{natbib}
\usepackage{graphicx}
\usepackage{xspace}
\usepackage{amsmath}

\newcommand{\msun}{$M_{\odot}$\xspace}

\newcommand{\rsun}{$R_{\odot}$\xspace}

\newcommand{\mearth}{$M_\earth$\xspace}

\newcommand{\rearth}{$R_\earth$\xspace}

\newcommand{\feh}{\ensuremath{[\mbox{Fe}/\mbox{H}]}\xspace}

\newcommand{\Rmnum}[1]{\expandafter\@slowromancap\romannumeral #1@}

\newcommand{\microns}{\ensuremath{\mu\mathrm{m}}\xspace}

\newcommand{\mstar}{\ensuremath{M_{\star}}\xspace}
\newcommand{\rstar}{\ensuremath{R_{\star}}\xspace}

\newcommand{\fe}{$[$Fe/H$]$\xspace}
\newcommand{\teff}{\ensuremath{T_{\mathrm{eff}}}\xspace}
\newcommand{\logg}{\ensuremath{\log g}\xspace}

\newcommand{\gcc}{g\,cm$^{-3}$\xspace}

\newcommand{\rhostar}{\ensuremath{\rho_{\star}}\xspace}

\newcommand{\Teq}{$T_{\mathrm{eq}}$\xspace}

\newcommand{\corot}{\textit{CoRoT}\xspace}
\newcommand{\kepler}{\textit{Kepler}\xspace}
\newcommand{\ktwo}{\textit{K2}\xspace}
\newcommand{\spitzer}{\textit{Spitzer}\xspace}
\newcommand{\tess}{\textit{TESS}\xspace}
\newcommand{\jwst}{\textit{JWST}\xspace}
\newcommand{\cheops}{\textit{CHEOPS}\xspace}

\newcommand{\gaia}{\textit{Gaia}\xspace}

\newcommand{\ktwophot}{\texttt{k2phot}\xspace}
\newcommand{\terra}{\texttt{TERRA}\xspace}
\newcommand{\specmatchsyn}{\texttt{SpecMatch-syn}\xspace}
\newcommand{\specmatchemp}{\texttt{SpecMatch-emp}\xspace}
\newcommand{\vespa}{\texttt{vespa}\xspace}
\newcommand{\celerite}{\texttt{celerite}\xspace}
\newcommand{\radvel}{\texttt{RadVel}\xspace}

\begin{document}

\title{{\it Spitzer} transit follow-up of planet candidates from the \ktwo mission}
\shorttitle{{\it Spitzer} follow-up of \ktwo}
\shortauthors{Livingston et al.}

\author{John H.\ Livingston}
\affiliation{Department of Astronomy, University of Tokyo, 7-3-1 Hongo, Bunkyo-ku, Tokyo 113-0033, Japan}
\affiliation{JSPS Fellow}
\affiliation{\href{mailto:livingston@astron.s.u-tokyo.ac.jp}{{\tt livingston@astron.s.u-tokyo.ac.jp}}}

\author{Ian J.\ M.\ Crossfield}
\affiliation{Department of Physics, Massachusetts Institute of Technology, Cambridge, MA 02139, USA}

\author{Michael W.\ Werner}
\affiliation{Jet Propulsion Laboratory, 4800 Oak Grove Dr, Pasadena, CA 91109, USA}

\author{Varoujan Gorjian}
\affiliation{Jet Propulsion Laboratory, 4800 Oak Grove Dr, Pasadena, CA 91109, USA}

\author{Erik A.\ Petigura}
\affiliation{Department of Astronomy, California Institute of Technology, 1200 E California Blvd, Pasadena, CA 91125, USA}
\affiliation{Hubble Fellow}

\author{David R.\ Ciardi}
\affiliation{Caltech/IPAC-NASA Exoplanet Science Institute, 1200 E California Blvd, Pasadena, CA 91125, USA}

\author{Courtney D.\ Dressing}
\affiliation{Astronomy Department, University of California, Berkeley, CA 94720, USA}

\author{Benjamin J.\ Fulton}
\affiliation{Geological and Planetary Sciences, California Institute of Technology, 1200 E California Blvd, Pasadena, CA 91125, USA}

\author{Teruyuki Hirano}
\affiliation{Department of Earth and Planetary Sciences, Tokyo Institute of Technology, 2-12-1 Ookayama, Meguro-ku, Tokyo 152-8551, Japan}

\author{Joshua E.\ Schlieder}
\affiliation{NASA Goddard Space Flight Center, 8800 Greenbelt Rd, Greenbelt, MD 20771, USA}

\author{Evan Sinukoff}
\affiliation{Institute for Astronomy, University of Hawai'i at M\={a}noa, Honolulu, HI, USA}
\affiliation{Department of Astronomy, California Institute of Technology, 1200 E California Blvd, Pasadena, CA 91125, USA}

\author{Molly Kosiarek}
\affiliation{Department of Astronomy, University of California, Santa Cruz, Santa Cruz, CA, USA}
\affiliation{NSF Graduate Research Fellow}

\author{Rachel Akeson}
\affiliation{Caltech/IPAC-NASA Exoplanet Science Institute, 1200 E California Blvd, Pasadena, CA 91125, USA}

\author{Charles A.\ Beichman}
\affiliation{Caltech/IPAC-NASA Exoplanet Science Institute, 1200 E California Blvd, Pasadena, CA 91125, USA}

\author{Bj\"orn Benneke}
\affiliation{D\'epartement de Physique, Universit\'e de Montr\'eal, Montr\'eal, Quebec, H3C 3J7, Canada}

\author{Jessie L.\ Christiansen}
\affiliation{Caltech/IPAC-NASA Exoplanet Science Institute, 1200 E California Blvd, Pasadena, CA 91125, USA}

\author{Bradley M.\ S.\ Hansen}
\affiliation{Mani L.\ Bhaumik Institute for Theoretical Physics, Department of Physics \& Astronomy,
University of California Los Angeles, Los Angeles, CA 90095, USA}

\author{Andrew W.\ Howard}
\affiliation{Department of Astronomy, California Institute of Technology, 1200 E California Blvd, Pasadena, CA 91125, USA}

\author{Howard Isaacson}
\affiliation{Astronomy Department, University of California, Berkeley, CA 94720, USA}

\author{Heather A. Knutson}
\affiliation{Geological and Planetary Sciences, California Institute of Technology, 1200 E California Blvd, Pasadena, CA 91125, USA}

\author{Jessica Krick}
\affiliation{Caltech/IPAC-NASA Exoplanet Science Institute, 1200 E California Blvd, Pasadena, CA 91125, USA}

\author{Arturo O.\ Martinez}
\affiliation{Department of Physics and Astronomy, Georgia State University, 25 Park Place NE, Atlanta, GA 30303, USA}

\author{Bun'ei Sato}
\affiliation{Department of Earth and Planetary Sciences, Tokyo Institute of Technology, 2-12-1 Ookayama, Meguro-ku, Tokyo 152-8551, Japan}

\author{Motohide Tamura}
\affiliation{Department of Astronomy, University of Tokyo, 7-3-1 Hongo, Bunkyo-ku, Tokyo 113-0033, Japan}
\affiliation{Astrobiology Center, NINS, 2-21-1 Osawa, Mitaka, Tokyo 181-8588, Japan}
\affiliation{National Astronomical Observatory of Japan, NINS, 2-21-1 Osawa, Mitaka, Tokyo 181-8588, Japan}

\begin{abstract}
We present precision 4.5 \microns \spitzer transit photometry of eight planet candidates discovered by the \ktwo mission: K2-52\,b, K2-53\,b, EPIC\,205084841.01, K2-289\,b, K2-174\,b, K2-87\,b, K2-90\,b, and K2-124\,b. The sample includes four sub-Neptunes and two sub-Saturns, with radii between 2.6 and 18 \rearth, and equilibrium temperatures between 440 and 2000 K. In this paper we identify several targets of potential interest for future characterization studies, demonstrate the utility of transit follow-up observations for planet validation and ephemeris refinement, and present new imaging and spectroscopy data. Our simultaneous analysis of the \ktwo and \spitzer light curves yields improved estimates of the planet radii, and multi-wavelength information which help validate their planetary nature, including the previously un-validated candidate EPIC\,205686202.01 (K2-289\,b). Our \spitzer observations yield an order of magnitude increase in ephemeris precision, thus paving the way for efficient future study of these interesting systems by reducing the typical transit timing uncertainty in mid-2021 from several hours to a dozen or so minutes. K2-53\,b, K2-289\,b, K2-174\,b, K2-87\,b, and K2-90\,b are promising radial velocity (RV) targets given the performance of spectrographs available today or in development, and the M3V star K2-124 hosts a temperate sub-Neptune that is potentially a good target for both RV and atmospheric characterization studies.
\end{abstract}


\section{Introduction}
The NASA \ktwo mission \citep{2014PASP..126..398H} has extended the legacy of \kepler by discovering transiting exoplanets and candidate planets at a rate of hundreds per year. In contrast to \kepler, \ktwo surveyed a wider sky area at the cost of shorter time baseline per field, which has enabled the discovery of planets orbiting brighter stars. In addition to monitoring a greater number of bright stars than \kepler, \ktwo monitored more low mass stars than \kepler, partly as a result of its community-driven target selection. The result is that planets detected by \ktwo are generally more amenable to follow-up. To date, \ktwo has significantly enhanced the number of known small planets orbiting brighter stars than those surveyed by \kepler (\citealt{2015ApJ...806..215F}; \citealt{2015ApJ...809...25M}; \citealt{2016ApJS..222...14V}; \citealt[][hereafter Cr16]{2016ApJS..226....7C}; \citealt{2018AJ....155..136M}; \citealt{2018AJ....156...78L, 2018arXiv181004074L}), as well as discovering planets in cluster environments \citep{2016AJ....152..223O, 2017AJ....153..177P, 2016AJ....151..112D, 2016ApJ...818...46M, 2017AJ....153...64M, 2017MNRAS.464..850G, 2018AJ....155....4M, 2018AJ....155..115L, 2018AJ....155...10C, 2018AJ....156..195R, 2018arXiv180901968L}, including a 5--10 Myr planet in the Upper Scorpius star forming region \citep{2016Natur.534..658D, 2016AJ....152...61M}. \ktwo planets will be available for follow-up studies with the James Webb Space Telescope \citep[\jwst;][]{2006SSRv..123..485G} contemporaneously with the planets expected to be found by the Transiting Exoplanet Survey Satellite \citep[\tess;][]{2015JATIS...1a4003R}.

Our focus with \spitzer \citep{2004ApJS..154....1W} is on small planets orbiting K and M dwarfs discovered by \ktwo, which could potentially be good targets for future radial velocity (RV) or atmospheric characterization studies. We have been conducting \spitzer transit observations of planet candidates and using these data to refine estimates of their orbital and physical parameters. As \ktwo observes each field for approximately 80 days each, our observations play a critical role in refining the ephemerides due to the long time baseline they provide (typically 6--12 months longer). These results are part of an ongoing program, data from which has been used to ensure the feasibility of future study of \ktwo planets by \citet{2016ApJ...822...39B}, \citet{2017ApJ...834..187B}, \citet{2018arXiv180110177C}, \citet{2018AJ....156...70D}, and Hardegree-Ullman, K. et al. (in preparation).

In this paper we validate the planet K2-289\,b, identify several targets of potential interest for future characterization studies, and demonstrate the utility of transit follow-up observations for ephemeris refinement. When done with a smaller beam, i.e. with \spitzer or \cheops \citep{2014SPIE.9143E..2JF}, such follow-up will prove especially useful in the validation of planet candidates identified by \tess, which will frequently encounter stellar blends due to the $\sim$21\arcsec\, pixel scale of its detectors. In \autoref{sec:observations} we describe our observations, including \ktwo and \spitzer photometry, high resolution imaging and spectroscopy, and literature data. In \autoref{sec:analysis} we describe the analysis methods used to measure host star and planet properties from these data, as well as our planet validation approach. In \autoref{sec:discussion} we present the results of our analyses and discuss the potential for future characterization studies, concluding with a summary in \autoref{sec:summary}.

\section{Observations}
\label{sec:observations}

\subsection{K2 photometry}
\label{sec:observations-k2}

The basis for this work is the initial identification of planet candidates in \ktwo light curves. This process is described in Cr16 and \citet{2018AJ....155...21P}, but we briefly summarize it here. We use \ktwophot\footnote{\url{https://github.com/petigura/k2phot}} to correct the instrumental systematics induced by the roll of the \kepler spacecraft. The resulting corrected light curves are publicly available on the community portal ExoFOP\footnote{\url{https://exofop.ipac.caltech.edu}}. We use \terra\footnote{\url{https://github.com/petigura/terra}} to search these light curves for transit signals, and the resulting candidates are then vetted by eye to eliminate instrumental or astrophysical false positives. During this process we also assess the utility of conducting follow-up transit observations with \spitzer. The planets we analyze here were all deemed interesting targets for \spitzer because they were relatively small, temperate, and/or orbit late-type host stars. All of the planets and candidates in this work were previously published by Cr16, with the exception of K2-124\,b, which was observed in \ktwo Campaign 5 and subsequently discovered by our team \citep{2017AJ....154..207D, 2018AJ....155...21P, 2018arXiv181004074L}.

\subsection{Spitzer photometry}
\label{sec:observations-spitzer}

\spitzer presents several advantages over ground-based transit follow-up observations: its position in space enables precise photometry unaffected by the Earth's atmosphere; its Earth-trailing orbit frees it from the scheduling constraints imposed by the day/night cycle on Earth; the diminished effects of limb darkening in the infrared enable precise estimation of transit model parameters; the 4.5 \microns Infrared Array Camera \citep[IRAC,][]{1998SPIE.3354.1024F} bandpass (in conjunction with the \kepler bandpass) provides a relatively broad wavelength baseline which facilitates planet validation. In addition, our high-cadence \spitzer observations provide better sampling of the transit than the 30 minute cadence of \ktwo.

We conducted our transit observations using the IRAC 4.5 \microns channel as part of \spitzer cycle 11 GO program 11026 (P.I. Werner). We chose integration times between 2--30 seconds to keep the detector in the linear regime and minimize downlink bandwidth. Target acquisition places the stars on the ``sweet spot'' of the detector, which has been well-characterized for the purpose of precise time-series photometry \citep{2012SPIE.8442E..1YI}, and falls within the region of the detector accessible in sub-array mode (used for observing bright stars). Following the guidelines for high precision \spitzer photometry \citep{2012SPIE.8448E..1IG}, we performed $\sim 30$ minutes of integrations on an empty field before each transit observation, which can help mitigate systematics induced by thermal settling of the spacecraft. See table \autoref{tab:log} for details of the observations.

\begin{deluxetable}{lcccc}
\tabletypesize{\scriptsize}
\tablecaption{Spitzer observing log \label{tab:log}}
\tablehead{EPIC & Name & Int. Time & Duration & Start Date \\
     &  & [sec] & [hr] & [JD] }
\startdata
203776696 & K2-52 & 30 & 10.6 & 2457337.6776 \\
204890128 & K2-53 & 6 & 8.1 & 2457347.8974 \\
205084841 & & 30 & 7.8 & 2457334.7969 \\
205686202 & K2-289 & 12 & 8.9 & 2457554.5479 \\
210558622 & K2-174 & 2$^\dagger$ & 13.5 & 2457533.4483 \\
210731500 & K2-87 & 30 & 11.8 & 2457519.4901 \\
210968143 & K2-90 & 12 & 10.4 & 2457531.5406 \\
212154564 & K2-124 & 30 & 8.3 & 2457590.9404 \\
\enddata
\tablecomments{$^\dagger$Due to the brightness of K2-174, the observations were conducted in sub-array mode to accommodate shorter integrations and limit the bandwidth required for the data downlink to Earth.}
\end{deluxetable}

\subsection{High resolution imaging}
\label{sec:observations-imaging}

\begin{figure}
\centering
\includegraphics[width=0.49\textwidth,trim={0cm 0 0 0cm}]{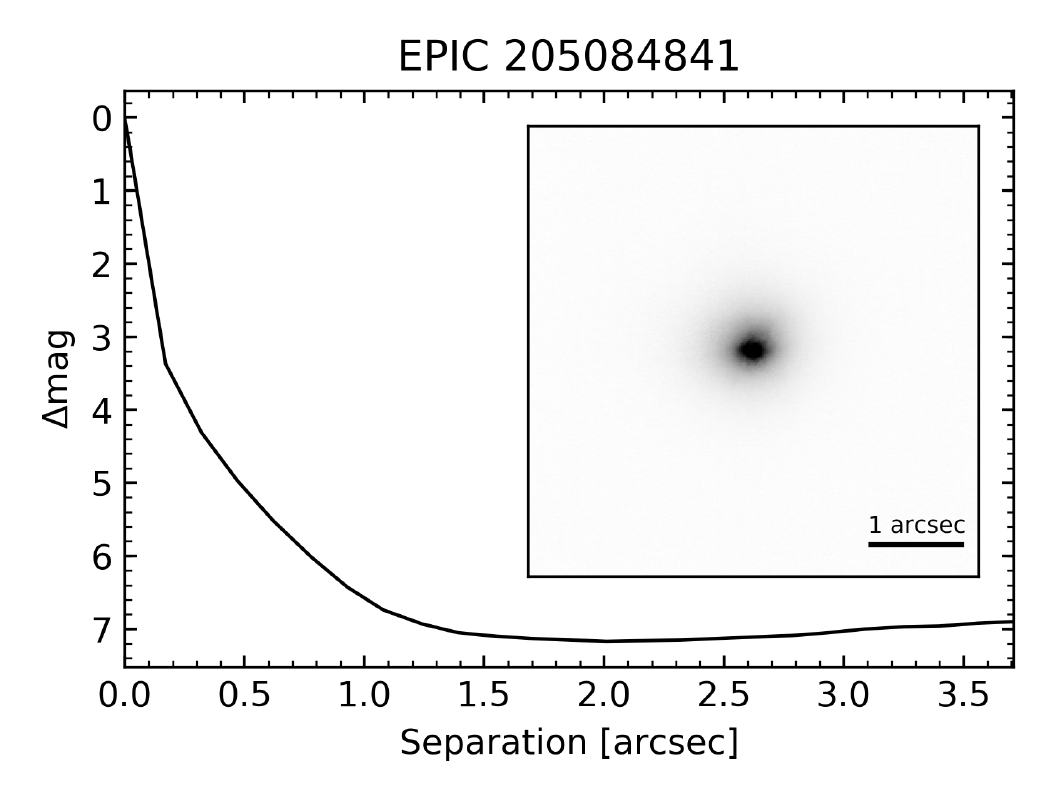}
\includegraphics[width=0.49\textwidth,trim={0cm 0 0 0cm}]{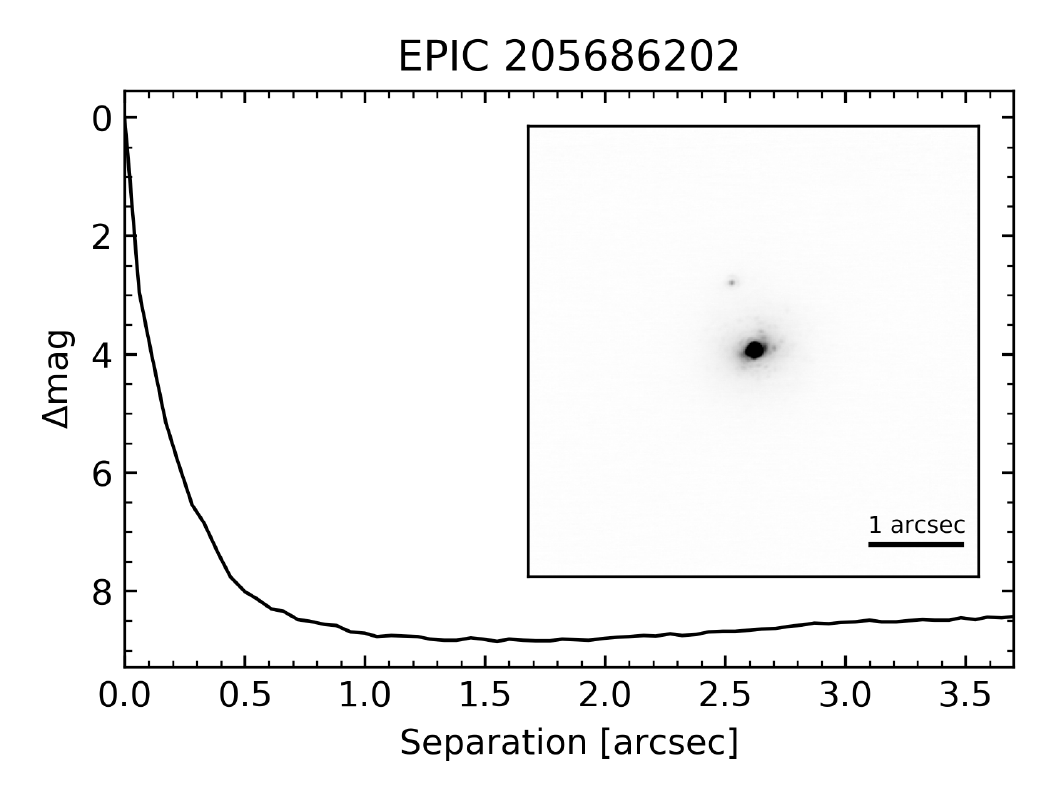}
\caption{Keck/NIRC2 $K$ band AO imaging of EPIC\,205084841 and EPIC\,205686202 (K2-289), and their resulting contrast curves.}
\label{fig:ao}
\end{figure}

High resolution imaging is important for detecting stellar companions and constraining the probability of chance alignments with background sources within the \ktwo and \spitzer photometric apertures, and thus plays a critical role in assessing the false positive probability (FPP) of a planet candidate. In this work, we utilize imaging previously published by Cr16 along with additional AO imaging from Keck/NIRC2 \citep{2014SPIE.9148E..2BW}. Cr16 did not obtain imaging of EPIC\,205084841, so we used NIRC2 in natural guide star mode to observe the star in $K$ band on UT July 9, 2017. Using the image reduction and analysis methods described in Cr16, we find the star to be single and rule out companions above the contrast curve shown in \autoref{fig:ao} at the 5$\sigma$ confidence level. We also utilize a $J$ band image of K2-289, which was obtained with NIRC2 on UT April 1, 2015 and made available on ExoFOP, but was not published in Cr16. As in the $K$ band image reported by Cr16, the companion is clearly detected in $J$ band, thus providing useful color information (see \autoref{sec:discussion-validation}).

\subsection{Spectroscopy}
\label{sec:observations-spectroscopy}

Our team has primarily used Keck/HIRES \citep{1994SPIE.2198..362V} to obtain high resolution spectra of candidate host stars, enabling the measurement of more robust planet properties as well as detecting (or ruling out) double-lined spectroscopic binaries (see Cr16 and \citet{2018AJ....155...21P} for more details). We use \specmatchsyn to measure precise stellar parameters from the spectra of stars hotter than $\sim$4200 K, which matches spectra to an interpolated grid of models from \citet{2005A&A...443..735C}. For cooler stars, we use \specmatchemp, which matches spectra to a spectral library of 404 standard stars \citep{2017yCat..18360077Y}.

We also conducted spectroscopic observations of K2-289 using the High Dispersion Spectrograph (HDS) mounted on the Subaru 8.2 m telescope between UT 2016 April 29 and May 2. We employed the standard I2a setup, which covers $4940-7590\mathrm{\AA}$, and the image slicer \#2, achieving a spectral resolution of $R\sim 80,000$. For the RV measurements, we used the iodine (I$_2$) cell on three consecutive nights. We also obtained the stellar spectrum without the I$_2$ cell for the template of RV measurements. The HDS data were reduced using standard IRAF routines, by which we extracted one-dimensional, wavelength-calibrated spectra of K2-289. We then put those spectra through a RV analysis pipeline \citep{2002PASJ...54..873S, 2012PASJ...64...97S}, which does a forward modeling of each observed spectrum to measure the RV relative to the template. \autoref{tab:rvs} lists the extracted relative RVs and their internal errors.

\begin{deluxetable}{lcc}
\tabletypesize{\scriptsize}
\tablecaption{Radial velocities of K2-289 obtained with Subaru/HDS. \label{tab:rvs}}
\tablehead{BJD	&	RV (relative)	&	Error	\\
	&	[m\,s$^{-1}$]	&	[m\,s$^{-1}$]}
\startdata
2457509.063075 & -5.95 & 23.95 \\
2457510.065150 & 21.30 & 21.79 \\
2457510.086821 & 62.27 & 22.18 \\
2457511.067174 & -10.08 & 21.78 \\
2457511.088864 & -16.61 & 24.39 \\
\enddata
\end{deluxetable}

\section{Analysis}
\label{sec:analysis}

\subsection{\ktwo light curves}
\label{sec:analysis-k2}

As part of our team's large-scale transit search of the \ktwo data (\autoref{sec:observations-k2}), we correct systematics induced by the coupling of spacecraft motion and intra-pixel gain variations with \ktwophot, which uses a Gaussian Process model \citep[GP;][]{Rasmussen:2005:GPM:1162254}. The resulting light curves are essentially free of the large amplitude position-dependent flux variations characteristic of raw \ktwo photometry, but stellar variability and potentially residual systematic trends must be accounted for to measure precise transit properties. A common approach is to first model and remove out-of-transit variability using various methods, such as a median filter, polynomial, or spline model. These approaches are simple and fast, and usually do not significantly impact the results for stars with low levels of variability and residual systematics. In order to minimize the potential for biased parameter estimates while employing a uniform framework for a range of light curve behaviors, we use the \celerite GP framework with a Matern-3/2 kernel and take an approach similar to the ``type-II'' maximum likelihood estimation (MLE) described in \citet{2012MNRAS.419.2683G}.

The GP is first trained on the out-of-transit light curve using {\tt scipy.optimize}, and then the full light curve is analyzed with the same GP in conjunction with the transit model (\autoref{sec:analysis-transitmodel}). We first use the ``L-BFGS-B'' method \citep{53712fe04a3448cfb8598b14afab59b3, Zhu:1997:ALF:279232.279236} to find the MLE value of the kernel hyperparameters given the out-of-transit light curve, using the GP likelihood and gradient in \celerite. During this stage, we perform iterative outlier rejection to minimize the possibility of biased kernel hyperparameters. Next we use the Nelder-Mead simplex algorithm \citep{doi:10.1093/comjnl/7.4.308} to fit the joint GP and transit model to the full light curve, initialized with the MLE kernel hyperparameters and an initial set of transit parameters from previous analyses. Fitting the GP hyperparameters simultaneously with the transit model helps ensure that we find an optimal noise model that is valid during and out of transit. We then restrict the data to 1-day windows centered on each transit and sample the posterior of the joint GP-transit model, by running {\tt emcee} for 500 steps initialized with the optimum found in the previous step. This brief sampling stage is especially important if the kernel hyperparameters are initially stuck in a local optimum, and can be used to ensure that the posteriors are unimodal and sharply peaked around these optimal values.

\begin{figure*}
    \centering
    \includegraphics[width=0.88\textwidth,trim={0cm 0 0.05cm 0}]{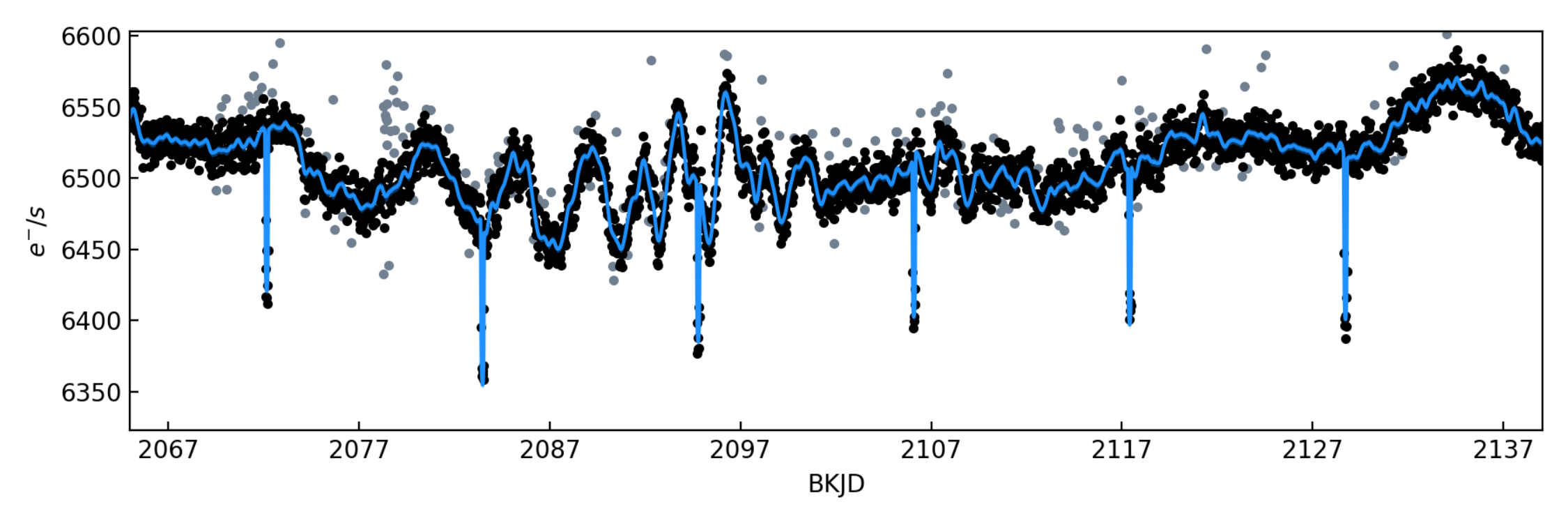}
    \includegraphics[width=0.88\textwidth,trim={0.1cm 0 0 0}]{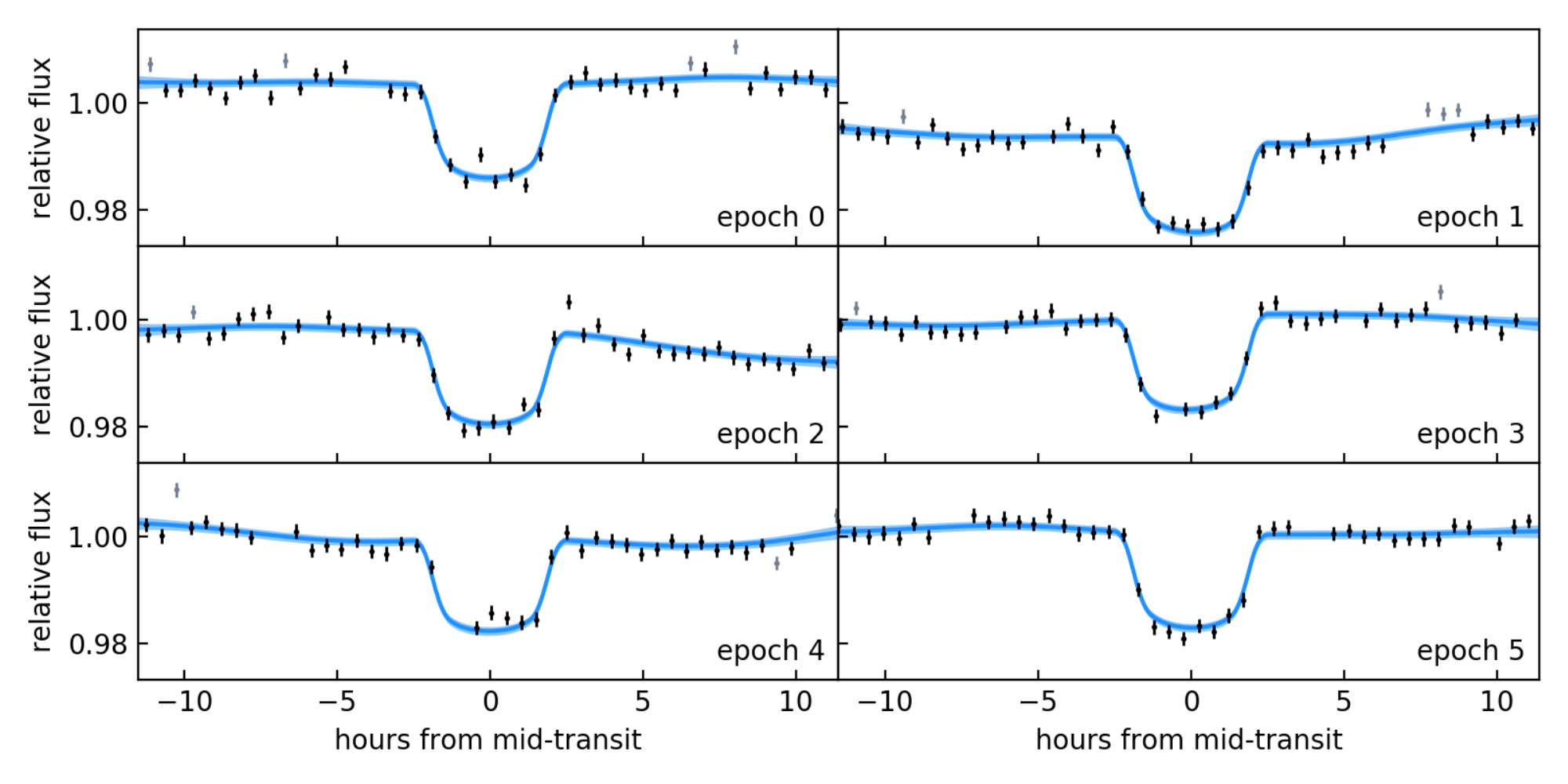}
    \caption{Top: \ktwo photometry of EPIC\,205084841 from \ktwophot with the best-fit GP model, and the GP-detrended light curve with best-fit transit model. Bottom: individual transits in the \ktwo light curve with the best-fit GP and transit models to illustrate the quality of the detrending in the top panel. In all panels the data are shown in black, outliers are shown in gray, models are shown in blue, and the shaded blue regions show the 2$\sigma$ uncertainty from the GP model.}
    \label{fig:k2-4841}
\end{figure*}

\begin{figure*}
    \centering
    \includegraphics[width=0.88\textwidth,trim={0cm 0 0.05cm 0}]{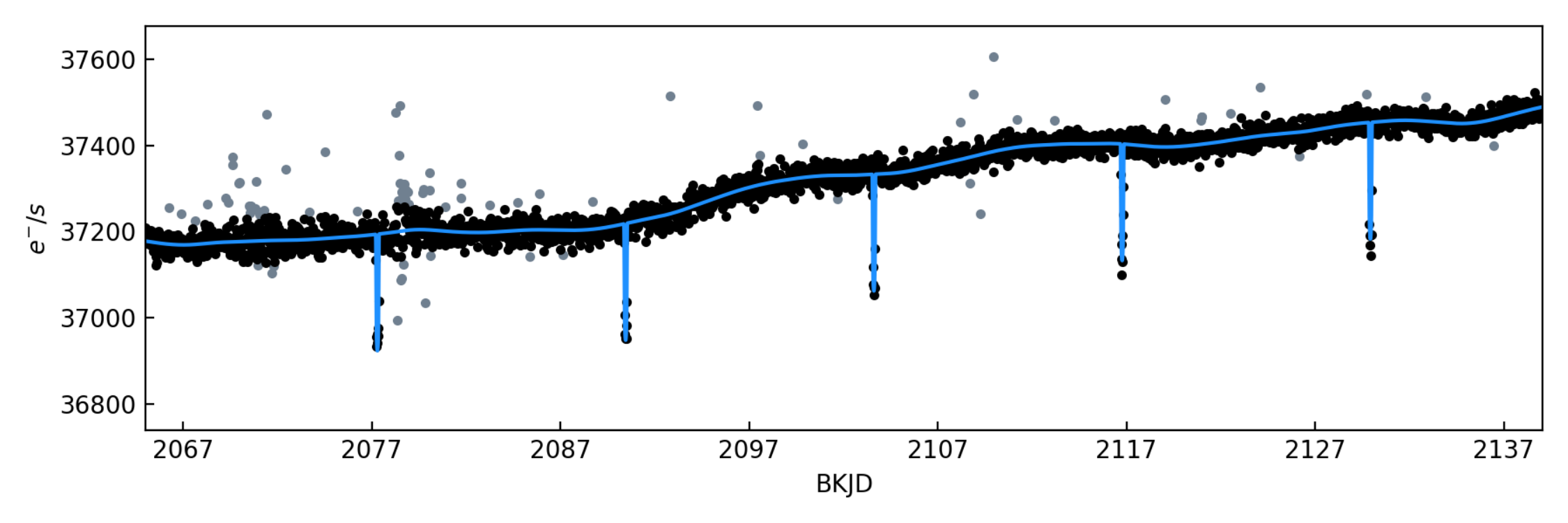}
    \includegraphics[width=0.87\textwidth,trim={0.1cm 0 0.1cm 0}]{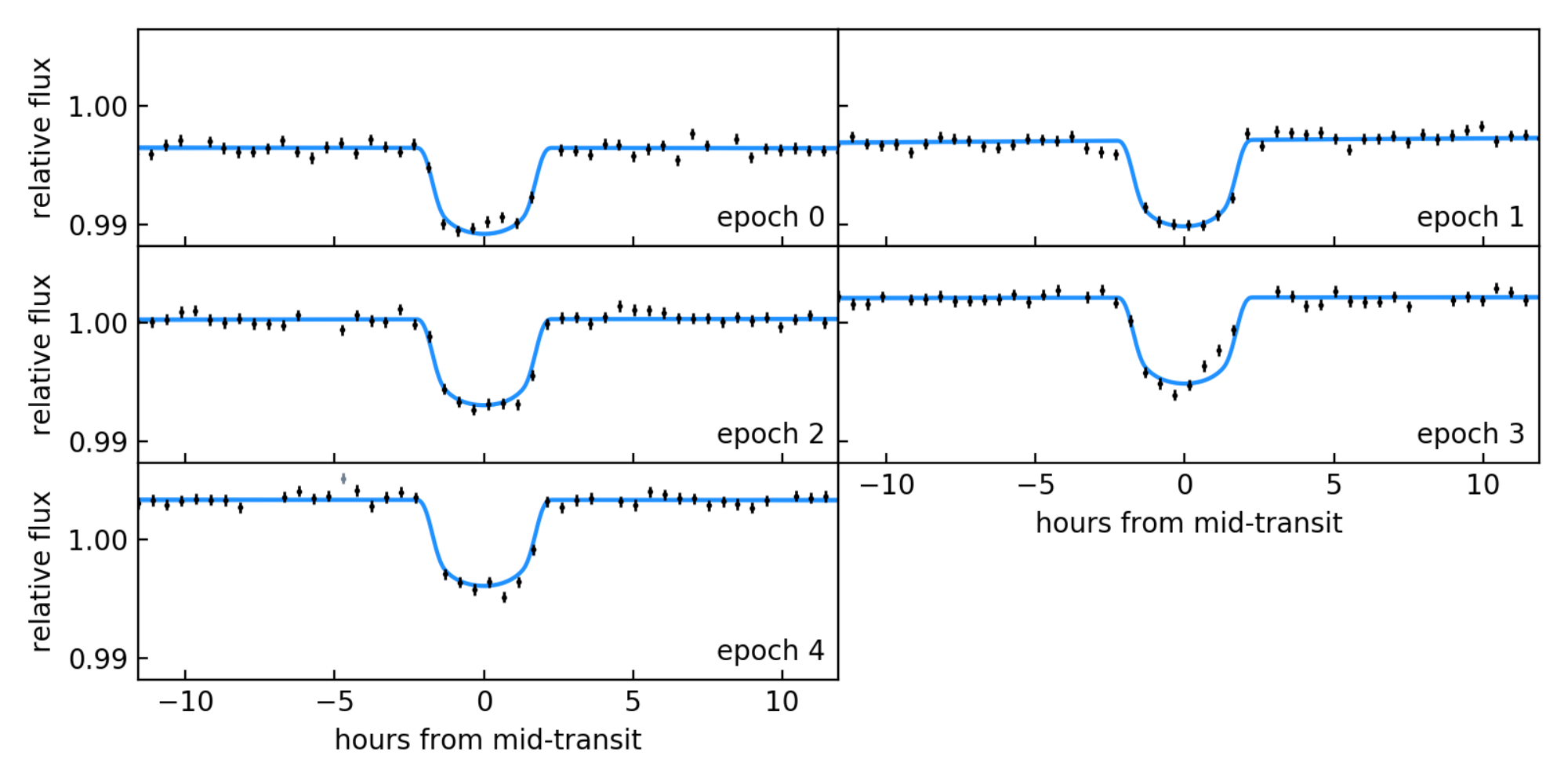}
    \caption{Same as \autoref{fig:k2-4841}, but for K2-289\,b. In-transit variability could be the result of spot-crossing, or simply residual systematics. An increase in the number of outliers can be seen near BKJD=2079 in this light curve as well as the one in \autoref{fig:k2-4841}, which suggests an instrumental origin.}
    \label{fig:k2-6202}
\end{figure*}

Finally, we fix the hyperparameters to their optima and proceed to run the sampler for 10,000 steps (see also \autoref{sec:analysis-transitmodel}). This approach retains the benefits of a flexible noise model while minimizing computational complexity. \autoref{fig:k2-4841} and \autoref{fig:k2-6202} show examples of these fits, with datapoints shown in gray if they were excluded during the iterative outlier rejection performed during the initial GP training stage.

\subsection{\spitzer light curve extraction}
\label{sec:analysis-spitzer-lightcurves}

We extract the \spitzer light curves following the approach taken by \citet{2012ApJ...754...22K} and \citet{2016ApJ...822...39B}. In brief, we compute aperture photometry using circular apertures centered on the host star, for a range of radii between 2.0 and 5.0 pixels, corresponding to 2.4--6.0\arcsec\ due to \spitzer's 1.2\arcsec\ pixel scale. We used a step size of 0.1 pixels from 2.0 to 3.0, and a step size of 0.5 from 3.0 and 5.0. An important component of precision photometry with \spitzer is the selection of an optimal aperture, due to the fact that significant levels of ``red" (correlated) noise present to a varying extent in each time series. Smaller radii tend to have less photon noise due to the decreased sky background in the aperture, while slightly larger radii can sometimes mitigate the inter- and intra-pixel gain variations which are responsible for the correlated noise. Ideally, an optimal radius minimizes both of these effects, although in practice it is common to attempt only to minimize correlated noise or the photon noise \citep[e.g.][]{2012ApJ...754...22K, 2013ApJ...766...95L, 2014A&A...572A..73L}.

A typical transit dataset contains significant time both in and out of transit, so we compute relevant noise metrics as a function of radius for different subsets, most of which are fully out-of-transit or between 2$^\mathrm{nd}$ and 3$^\mathrm{rd}$ contact (i.e. do not contain ingress or egress). Thus, for a given radius, the ensemble of these values are largely unaffected by the transit signal, and thus reflect only photon noise and systematic noise. To quantify the level of red noise, we compute $\beta$, the factor by which the standard deviation of the observed binned residuals deviate from the theoretical value \citep{2006MNRAS.373..231P, 2008ApJ...683.1076W}:
\begin{align}
\beta = \frac{\sigma_M}{\sigma_0} \sqrt{\frac{N(M-1)}{M}},
\end{align}
where $\sigma_M$ is the standard deviation of the binned residuals (in M bins), $\sigma_0$ is the standard deviation of the un-binned residuals, and N is the number of data points per bin. To ensure a robust estimate and focus on the timescales of red noise which could significantly impact transit parameter estimates, we compute the median $\beta$ value for bin widths between 5 and 40 minutes. We divide each flux time series into 10 equal-sized segments and compute both the standard deviation (i.e. overall noise level) and the $\beta$ value (i.e. red noise level) for each segment. The optimal aperture is then the one that minimizes each metric. We then compute the median of the optimal aperture radii for each metric over all segments. Finally, the aperture radius adopted for subsequent analysis is the mean of these two ``optimal'' radii; the selected aperture thus only ``approximately'' minimizes both metrics in cases where these two radii are not equal. We chose 10 segments as a tradeoff between having more robust statistics and having enough light curve in each segment to compute red noise on a range of different timescales. We find that the minimum red noise aperture is frequently consistent with the minimum standard deviation aperture to within a few tenths of a pixel, and the optimal radius is typically 2.2--2.4 pixels, which is consistent with the optimal apertures found in previous analyses of \spitzer transit data \citep[e.g.][]{2012ApJ...754...22K}, in which the residuals computed from the best-fit transit and systematics model are analyzed instead of the raw light curve.

In principle, aperture selection could be handled via Bayesian model selection (i.e. computing the Bayesian evidence), although any improvements might not be significant enough to justify the computational cost. Additionally, it may be fruitful to simultaneously estimate the white and red noise levels using a GP; a sufficient choice of kernel could more fully disentangle these two noise signals, as compared to the standard deviation and $\beta$ factor. We leave an investigation of these possibilities for a future work.

\subsection{\spitzer systematics model}
\label{sec:analysis-spitzer-systematics}

\begin{figure}
    \centering
    \includegraphics[width=0.45\textwidth,trim={1.25cm 0.25cm 0.75cm 0.25cm}]{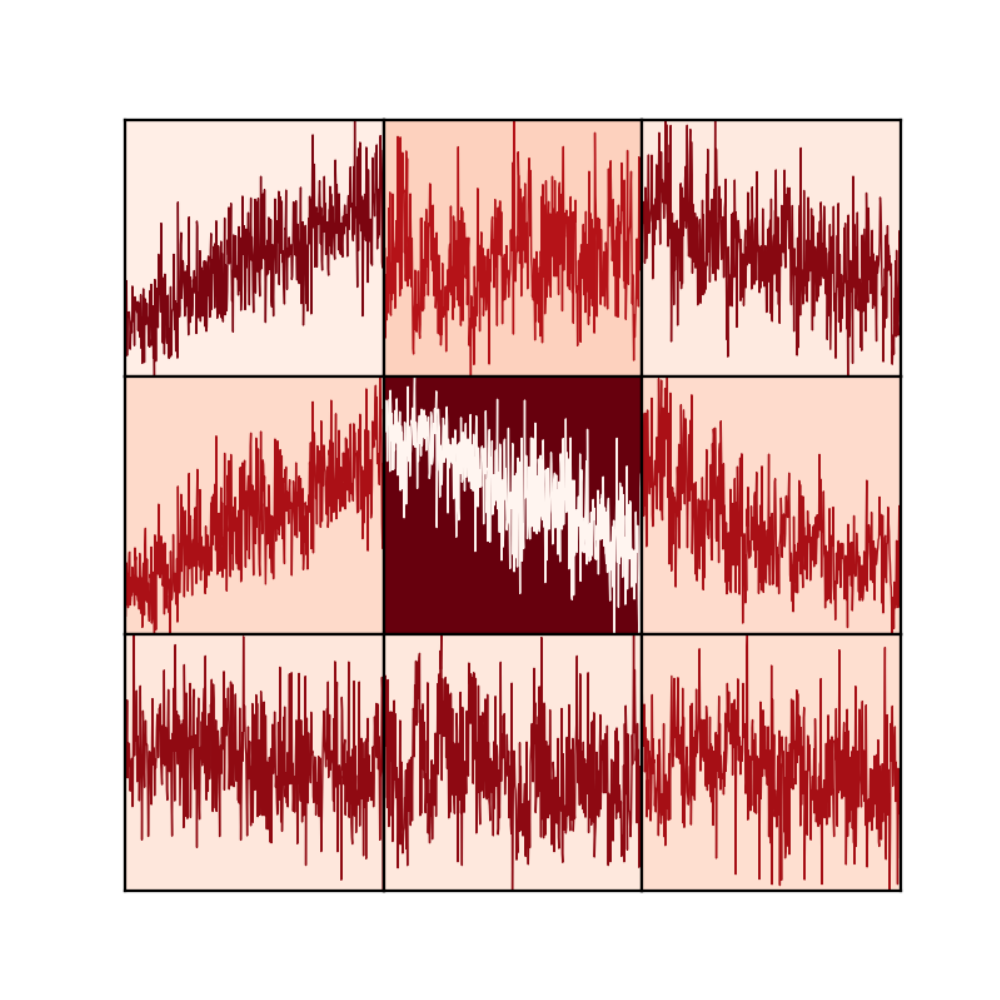}
    \includegraphics[width=0.45\textwidth,trim={1.25cm 0.25cm 0.75cm 0.25cm}]{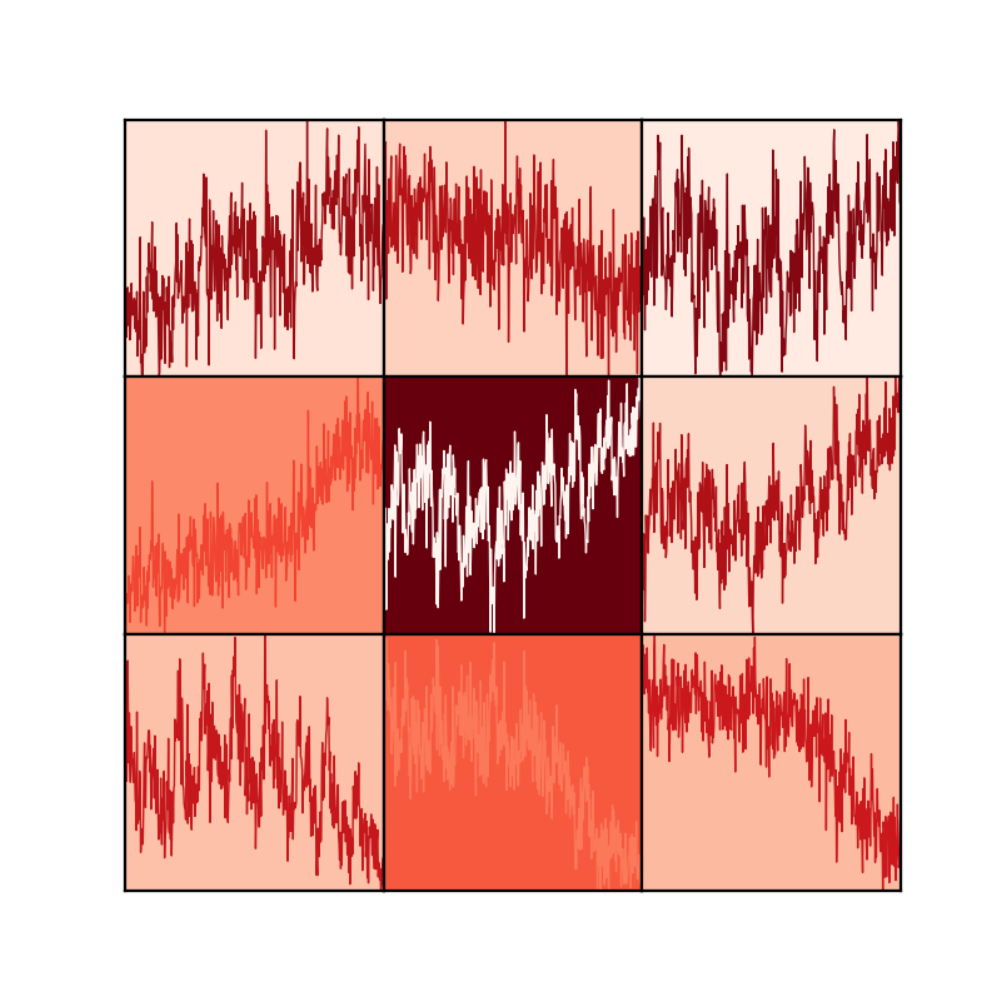}
    \caption{The normalized PLD pixel light curves of EPIC\,205084841 (top) and K2-289 (bottom), displayed in a grid corresponding to their location on the detector, with time on the x-axis. The color of the background of each cell illustrates the relative intensity of each pixel, where lighter colors correspond to higher intensity.}
    \label{fig:pld4841}
\end{figure}

\begin{figure*}
\centering
\includegraphics[width=0.9\textwidth,clip,trim={1cm 0.25cm 1.5cm 0.25cm}]{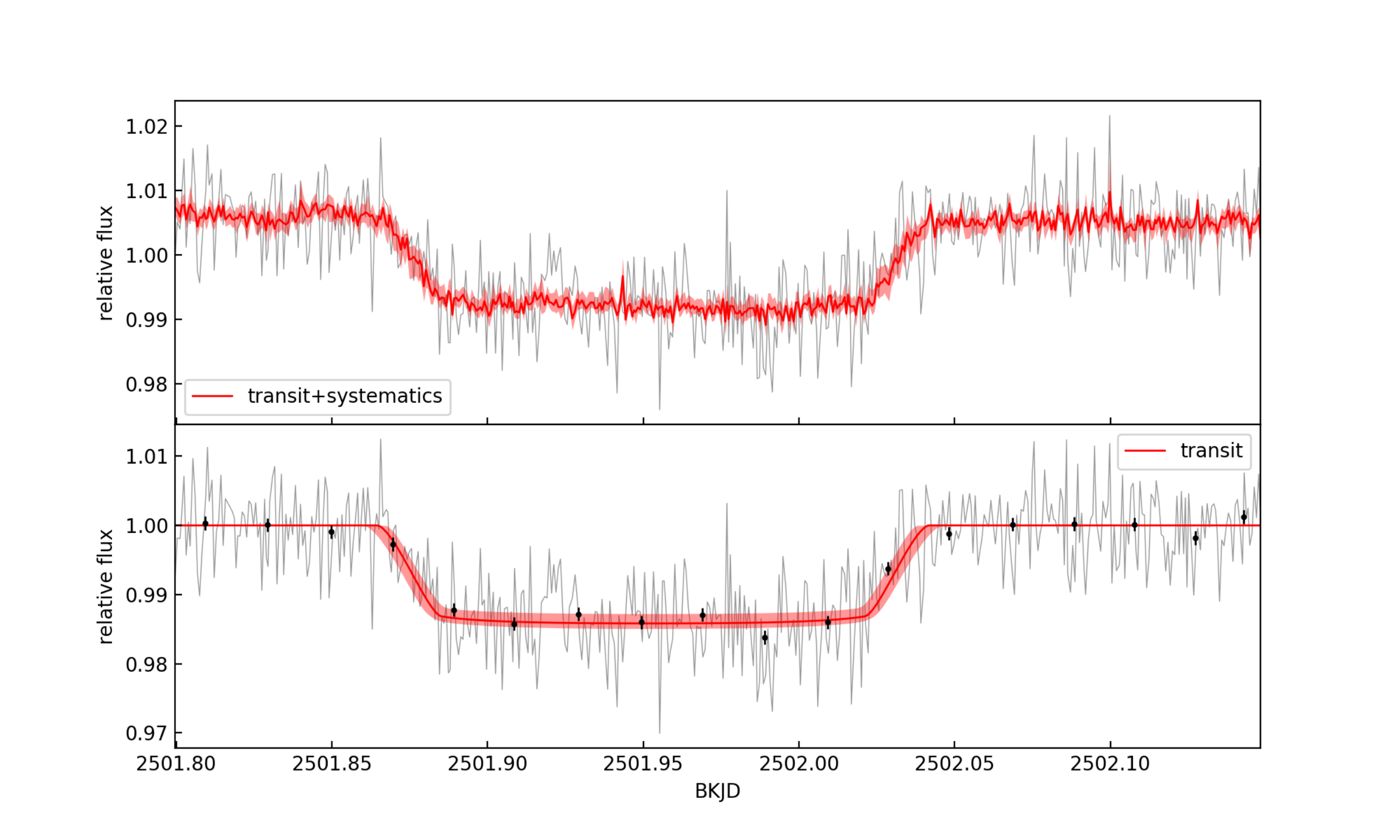}
\includegraphics[width=0.9\textwidth,clip,trim={1cm 0.25cm 1.5cm 0.25cm}]{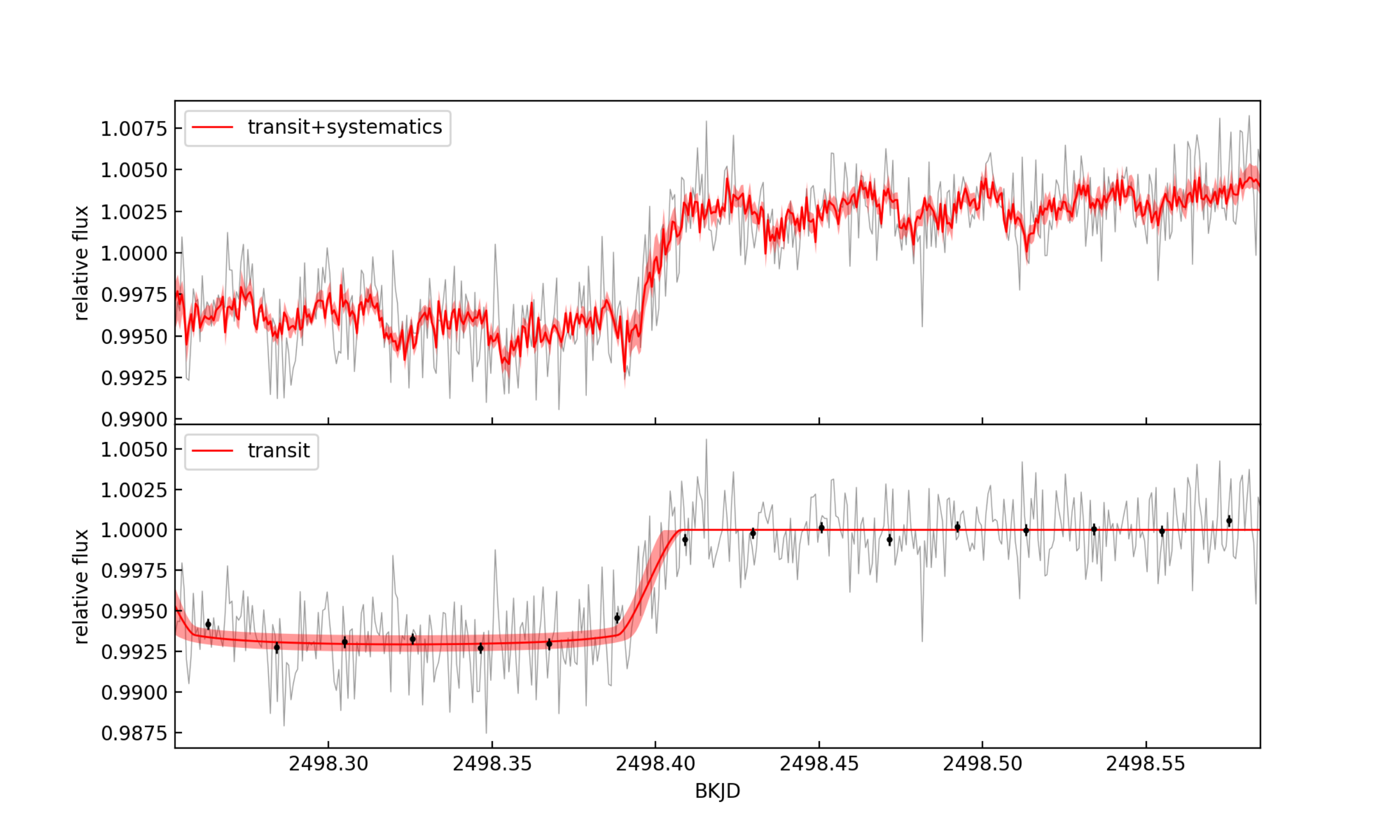}
\caption{Best fit to the \spitzer EPIC\,205084841.01 (top) and K2-289\,b (bottom) light curves. The top panels show the raw light curve with the best fitting model (transit + systematics) over-plotted, normalized to the median flux during the observations.  The lower panels show the data with the best-fitting systematics component removed and the best-fitting transit model over-plotted, normalized to unity out-of-transit flux. The 95\% credible regions from our MCMC analysis are shown as shaded regions. To aid comparison with the \ktwo data, the systematics-corrected photometry is also binned to the $\sim$30 min \ktwo observing cadence and shown as gray points, with error bars illustrating the Gaussian noise from the fit (scaled appropriately for the bin size).}
\label{fig:best4841-6202}
\end{figure*}

\begin{figure}
\centering
\includegraphics[width=0.49\textwidth,clip,trim={0cm 2cm 0.5cm 2.5cm}]{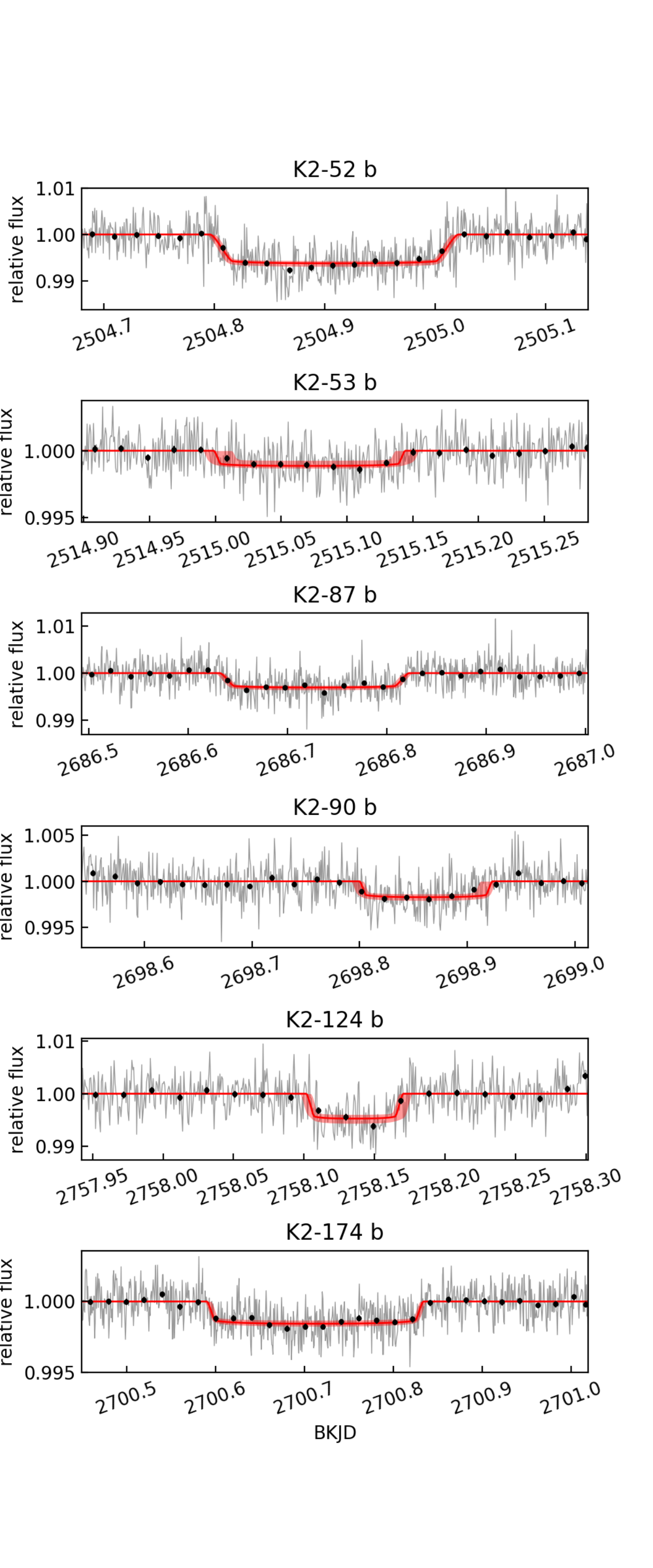}
\caption{Same as the lower panels in \autoref{fig:best4841-6202}, but for the remaining datasets.}
\label{fig:spitzer-grid}
\end{figure}

We model the systematics inherent to the \spitzer light curves using the pixel-level decorrelation (PLD) method \citep{2015ApJ...805..132D}. In comparisons between various methods used to correct \spitzer systematics \citep{2016AJ....152...44I}, PLD was among the top performers, displaying both high precision and repeatability. PLD uses a linear combination of (normalized) pixel light curves to model the effect of PSF motion on the detector coupled with intra-pixel gain variations, thus it does not require the calculation of centroids. The parametrization of the full model for the transit light curve, including PLD, is:
\begin{align}
\Delta S^t = \frac{\sum_{i=1}^9 c_i P_i^t}{\sum_{i=1}^9 P_i^t} + M_{tr}(\boldsymbol{\theta},t) + \varepsilon(\sigma),
\label{eqn:model}
\end{align}
where $M_{tr}$ is the transit model, the $c_i$ are the PLD coefficients, and $\varepsilon(\sigma)$ are zero-mean Gaussian errors with width $\sigma$. To form a valid set of basis vectors for the instrumental systematics component of the light curve, the astrophysical signal in each individual pixel light curve is removed by normalization (the sum in the denominator of \autoref{eqn:model}). We show an illustrative example of these normalized pixel light curves in \autoref{fig:pld4841}. In testing, we found that using a $3 \times 3$ pixel grid sufficiently captures the information content corresponding to the motion of the PSF on the detector (which is typically $\lesssim$ a few tenths of a pixel). However, \spitzer target acquisition occasionally misses the ``sweet spot'' pixel, which may yield datasets with more pronounced systematics that could benefit from using a larger pixel grid (such as $5 \times 5$). In \autoref{fig:best4841-6202} we plot the full model fit for two datasets (upper panels), as well as the data corrected by subtracting the best-fit PLD noise model (lower panels). In \autoref{fig:spitzer-grid} we plot the corrected data and transit models for the remaining \spitzer datasets.

\subsection{Transit fitting}
\label{sec:analysis-transitmodel}

To model the transits in both the \ktwo and \spitzer light curves, we use the analytic model of \citet{2002ApJ...580L.171M}, assuming a circular orbit and a quadratic limb darkening law, as implemented in {\tt batman}. The free parameters $\boldsymbol{\theta}$ of the transit model are the planet-to-star radius ratio $R_p/R_{\star}$, the scaled semi-major axis $a/R_{\star}$, mid transit time $T_0$, orbital period $P$, impact parameter $b\equiv a\cos(i)/R_\star$, and the modified quadratic limb darkening coefficients $q_1$ and $q_2$, which efficiently sample the space of physically allowed limb darkening coefficients using the triangular sampling method of \citet{2013MNRAS.435.2152K}. These transformed coefficients are computed directly from the quadratic limb darkening coefficients $u_1$ and $u_2$ using equations 17 and 18 of \citet{2013MNRAS.435.2152K}, reproduced here for convenience:

\begin{align}
q_1 &\equiv (u_1 + u_2)^2,\\
q_2 &\equiv \frac{u_1}{2(u_1 + u_2)}.
\label{eqn:qspace}
\end{align}

Following Cr16, we use Gaussian limb darkening priors in our light curve analysis, which we derive from the coefficients for the \kepler and IRAC bandpasses tabulated by \citet{2012yCat..35460014C}. We use a Monte Carlo approach in which we sample the stellar parameters (\teff, \logg, \fe) of each star and interpolate the tabulated coefficients at the sampled stellar values. Because we sample in $q$-space, we convert the tabulated values of $u_1$/$u_2$ to $q_1$/$q_2$ to enable using a Gaussian prior in $q$-space. We then use the mean and standard deviation of the sampled coefficients to define the Gaussian priors (see \autoref{sec:ld} for more details). For the final set of parameter estimates listed in \autoref{tab:params} we also impose a prior on the mean stellar density determined from our {\tt isochrones} analysis (see \autoref{sec:analysis-stellar}), which yields more precise parameter estimates by leveraging more information about the host star. However, we also perform a parallel set of identical analyses without this prior, which provides the opportunity to compare the density from our stellar characterization to the independent measurement of the mean stellar density from the light curve $\rho_{\star,LC}$ (see \autoref{eqn:rho}). We discuss this in detail in \autoref{sec:discussion-validation}.

For Markov Chain Monte Carlo (MCMC) parameter estimation, we use {\tt emcee}, a Python implementation of the affine-invariant ensemble sampler \citep{2010CAMCS...5...65G}. We performed an initial optimization with {\tt lmfit} and positioned 128 ``walkers'' in a Gaussian ball centered on the optimum. We then ran the sampler for 10,000 steps allowing it to evolve according to the MCMC.
Finally, we checked for convergence by visual inspection of the trace, discarding the first 5000 samples as ``burn-in,'' and computed the autocorrelation time of each parameter using the Python package {\tt acor}\footnote{\url{https://github.com/dfm/acor}} to ensure we had collected a sufficient number of independent samples after burn-in.

\subsection{Simultaneous \ktwo and \spitzer analysis}
\label{sec:analysis-simultaneous}

In this work we simultaneously model the \ktwo and \spitzer light curves of each target. This is motivated by several factors. First, the posterior distributions of some transit model parameters are often distinctly non-Gaussian (e.g. $a/R_{\star}$), so simply imposing Gaussian priors derived from previous analysis of the \ktwo light curves could bias the resulting fits to the \spitzer data. Secondly, we leverage the \ktwo data to model the \spitzer data, because we simultaneously fit the systematics and transit models. Some transit model parameters ($a/R_{\star}$ and $b$) are shared between \spitzer and \ktwo, while others are distinct. We fit $R_p/R_{\star}$ separately for the \kepler and \spitzer bandpasses to enable the detection of false positive scenarios.

Our procedure is as follows. We first fit the \ktwo data alone, and form Gaussian priors on $T_0$ and $P$ based on the mean and standard deviation of the posterior distributions from MCMC. These priors are then used in a fit to the \spitzer transit data alone, yielding an initial set of parameter estimates derived from the \spitzer data. We then fit the \ktwo and \spitzer data simultaneously, without priors on $T_0$ and $P$. This simultaneous fit is responsible for the improvement in the ephemeris estimates, for reasons discussed above, as well as more robust transit shape parameter estimates. We list the transit parameters in \autoref{tab:params}, and the ephemeris estimates in \autoref{tab:ephemeris}.

\begin{figure}
\centering
\includegraphics[width=0.45\textwidth,trim={0.5cm 0cm 1.5cm 0.25cm}]{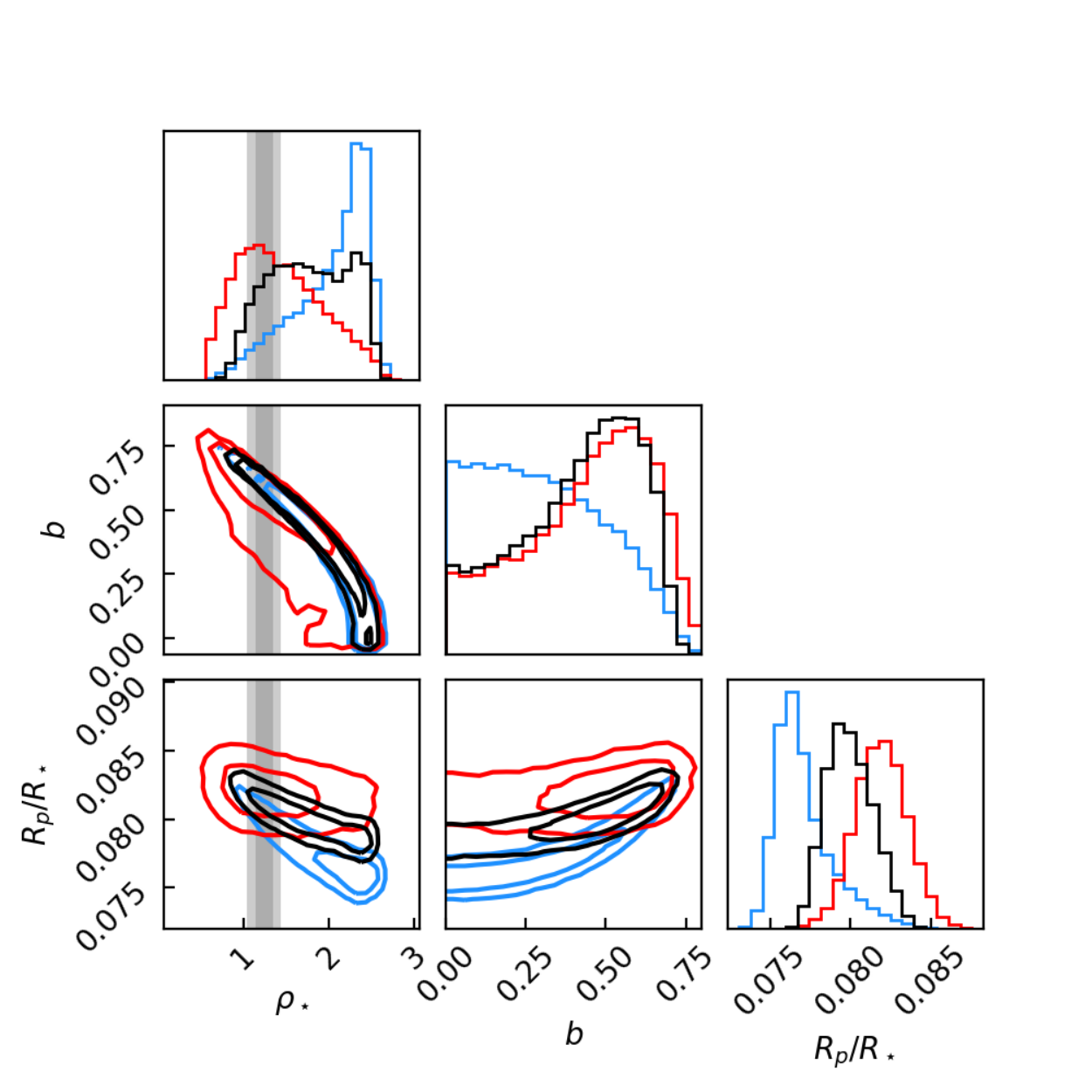}
\caption{Marginalized posterior distributions for key system parameters of K2-289\,b (without a prior on the mean stellar density), where the contours correspond to the 68\% and 95\% credible regions. The colors of the contours and histograms correspond to the dataset analyzed: blue is \ktwo-only, red is \spitzer-only, and black is simultaneous \ktwo-\spitzer. The vertical gray bands show the 68\% and 95\% credible regions for the stellar density from our {\tt isochrones} analysis.}
\label{fig:corner6202}
\end{figure}

An additional benefit to this approach is that it allows the high cadence and diminished limb darkening of our \spitzer light curves to yield improved constraints on transit parameters which are sensitive to ingress and egress (e.g. the impact parameter $b$). For example, even though our \spitzer observation (just barely) missed the ingress of K2-289\,b, (see \autoref{fig:best4841-6202}), the joint analysis of the \ktwo and \spitzer data yields better constraints on the transit geometry than either the \spitzer-only or \ktwo-only analyses. \autoref{fig:corner6202} shows several key posterior distributions for this system from \ktwo-only, \spitzer-only, and our joint analysis of both datasets, illustrating the improved constraints in transit geometry. The left-most panels in \autoref{fig:corner6202} show the stellar density from our {\tt isochrones} analysis as gray bands for comparison to the mean stellar density derived from \autoref{eqn:rho} and the transit fit posteriors (without a density prior). In this case the density estimated from the \ktwo data alone can be seen to be in mild disagreement with the {\tt isochrones} density, but the \spitzer data yield an improved density estimate in good agreement. The improvement in parameter estimates afforded by this simultaneous modeling approach can be thought of as a type of ``Bayesian shrinkage,'' in which the high cadence of \spitzer and high photometric precision of \ktwo work together to extract higher measurement precision from the data.

\begin{deluxetable*}{lcccccccccccc}
\tabletypesize{\scriptsize}
\rotate
\tablecaption{Transit parameter estimates \label{tab:params}}
\tablehead{Name & $a$ & $b$ & $i$ & $R_{p,K}$ & $R_{p,S}$ & $R_{p,C}$ & log($\sigma_K$) & log($\sigma_S$) & $T_{14}$ & $T_{23}$ & $\eta$ & $R_{p,\mathrm{max}}$ \\
 & [\rstar] & & [degrees] & [\rstar] & [\rstar] & [\rstar] & & & [days] & [days] & & [\rstar]}
\startdata
K2-52\,b     &   $5.2^{+0.2}_{-0.3}$     &  $0.32^{+0.14}_{-0.18}$     &  $86.46^{+2.04}_{-1.91}$     &  $0.072^{+0.001}_{-0.001}$     &  $0.078^{+0.002}_{-0.002}$     &  $0.075^{+0.001}_{-0.001}$     &  -7.01     &  -5.70     &  $0.222^{+0.003}_{-0.003}$     &  $0.187^{+0.002}_{-0.003}$     &  $0.84^{+0.01}_{-0.02}$     &  $0.085^{+0.013}_{-0.008}$  \\
K2-53\,b     &  $26.3^{+0.4}_{-0.4}$     &  $0.31^{+0.07}_{-0.10}$     &  $89.32^{+0.22}_{-0.16}$     &  $0.028^{+0.001}_{-0.001}$     &  $0.031^{+0.002}_{-0.002}$     &  $0.030^{+0.001}_{-0.001}$     &  -8.83     &  -6.61     &  $0.145^{+0.003}_{-0.003}$     &  $0.136^{+0.004}_{-0.003}$     &  $0.94^{+0.00}_{-0.00}$     &  $0.033^{+0.002}_{-0.002}$  \\
205084841.01     &  $17.8^{+0.5}_{-0.6}$     &  $0.62^{+0.03}_{-0.03}$     &  $87.99^{+0.16}_{-0.17}$     &  $0.127^{+0.002}_{-0.002}$     &  $0.117^{+0.002}_{-0.002}$     &  $0.122^{+0.002}_{-0.002}$     &  -6.36     &  -5.25     &  $0.189^{+0.003}_{-0.002}$     &  $0.125^{+0.003}_{-0.003}$     &  $0.66^{+0.02}_{-0.02}$     &  $0.203^{+0.016}_{-0.014}$  \\
K2-289\,b     &  $22.5^{+0.3}_{-0.3}$     &  $0.60^{+0.02}_{-0.02}$     &  $88.48^{+0.07}_{-0.07}$     &  $0.081^{+0.001}_{-0.001}$     &  $0.082^{+0.001}_{-0.002}$     &  $0.081^{+0.001}_{-0.001}$     &  -7.42     &  -6.24     &  $0.168^{+0.002}_{-0.002}$     &  $0.130^{+0.002}_{-0.002}$     &  $0.77^{+0.01}_{-0.01}$     &  $0.128^{+0.005}_{-0.005}$  \\
K2-174\,b     &  $26.6^{+0.2}_{-0.2}$     &  $0.02^{+0.03}_{-0.02}$     &  $89.95^{+0.04}_{-0.05}$     &  $0.033^{+0.000}_{-0.000}$     &  $0.038^{+0.001}_{-0.001}$     &  $0.036^{+0.001}_{-0.001}$     &  -9.25     &  -6.80     &  $0.242^{+0.002}_{-0.002}$     &  $0.225^{+0.002}_{-0.002}$     &  $0.93^{+0.00}_{-0.00}$     &  $0.036^{+0.001}_{-0.001}$  \\
K2-87\,b     &  $14.4^{+0.3}_{-0.3}$     &  $0.57^{+0.03}_{-0.03}$     &  $87.74^{+0.17}_{-0.17}$     &  $0.045^{+0.001}_{-0.001}$     &  $0.055^{+0.002}_{-0.002}$     &  $0.050^{+0.001}_{-0.001}$     &  -7.84     &  -5.94     &  $0.190^{+0.002}_{-0.002}$     &  $0.164^{+0.003}_{-0.003}$     &  $0.86^{+0.01}_{-0.01}$     &  $0.074^{+0.005}_{-0.004}$  \\
K2-90\,b      &  $33.4^{+0.3}_{-0.3}$     &  $0.44^{+0.05}_{-0.05}$     &  $89.24^{+0.09}_{-0.09}$     &  $0.036^{+0.001}_{-0.001}$     &  $0.041^{+0.002}_{-0.002}$     &  $0.038^{+0.001}_{-0.001}$     &  -8.33     &  -6.37     &  $0.123^{+0.003}_{-0.003}$     &  $0.112^{+0.003}_{-0.003}$     &  $0.91^{+0.01}_{-0.01}$     &  $0.048^{+0.004}_{-0.003}$  \\
K2-124\,b     &  $28.0^{+1.5}_{-1.7}$     &  $0.47^{+0.10}_{-0.14}$     &  $89.03^{+0.32}_{-0.27}$     &  $0.069^{+0.002}_{-0.002}$     &  $0.068^{+0.003}_{-0.003}$     &  $0.069^{+0.002}_{-0.002}$     &  -6.89     &  -5.73     &  $0.070^{+0.002}_{-0.002}$     &  $0.058^{+0.003}_{-0.003}$     &  $0.84^{+0.02}_{-0.03}$     &  $0.089^{+0.016}_{-0.013}$  \\

\enddata
\tablecomments{Transit parameters derived from the simultaneous analysis of the \ktwo and \spitzer data. The parameters $R_{p,K}$, $R_{p,S}$, and $R_{p,C}$ are the \ktwo, \spitzer, and combined values of the planet radius in units of the stellar radius. The best-fit logarithm of the Gaussian errors for the \ktwo and \spitzer time series are denoted by log($\sigma_K$) and log($\sigma_S$), respectively. The parameter $\eta$ is the transit shape defined in \autoref{eqn:shape}. The maximum planet radius allowed by the shape of the transit is denoted $R_{p,max}$, and defined in \autoref{eqn:rpmax}.}
\end{deluxetable*}

\subsection{Ephemerides}
\label{sec:analysis-ephemerides}

The addition of even a single follow-up transit measurement can result in significant refinement to the ephemeris of \ktwo planet candidates, due to the relatively small number of transits detected in each $\sim$80 day \ktwo observing campaign. The ephemerides from our simultaneous fit to the \ktwo and \spitzer data are often a highly significant improvement over the \ktwo-only ephemerides. To quantify this improvement, we compute the factor by which the precision on the estimate of the orbital period increases between the \ktwo-only and joint fits to the \ktwo and \spitzer data, which ranges from 8x to 13x (see \autoref{tab:ephemeris}).

The benefits of updating the ephemeris can be especially important in the context of planning future transit observations. Large uncertainties in the ephemeris estimates from the \ktwo light curves are common, due to the relatively short time span of each \ktwo observing campaign. The problem is particularly pronounced for planets with longer periods, which may transit only a small number of times in a given campaign. For those interested in studying the atmospheres of these planets, this complicates scheduling of future transit observations, due to the need to lengthen the observation window to ensure the transit is fully observed.

However, the addition of a single transit observation with \spitzer dramatically reduces the length of the necessary observation windows in the \jwst era. First, the longer time baseline yields an order of magnitude improvement in the precision of our orbital period estimates. Secondly, the new, combined transit mid-time results in smaller propagated timing uncertainties due to being closer in time to the planned transit observation than the \ktwo epoch. Our \spitzer observations reduce the typical uncertainty on the predicted transit time in mid-2021 from hours to minutes (see \autoref{tab:ephemeris}), enabling follow-up observations to be much more efficiently scheduled; even in 2025, the uncertainties will still be less than an hour.

\begin{deluxetable*}{lccccccc}
\tabletypesize{\scriptsize}
\tablecaption{Ephemerides \label{tab:ephemeris}}
\tablecolumns{8}
\tablehead{
\colhead{} &
\multicolumn{2}{c}{\ktwo-only} &
\multicolumn{2}{c}{\ktwo+\,\spitzer} &
\multicolumn{2}{c}{Mid-2021 Timing Uncertainty} &
\colhead{} \\
\cline{6-7}
\colhead{Name} &
\colhead{$P$} &
\colhead{$T_{0}$} &
\colhead{$P$} &
\colhead{$T_{0}$} &
\colhead{\ktwo-only} &
\colhead{\ktwo+\,\spitzer} &
\colhead{Improvement$^{\dagger}$} \\
\colhead{} &
\colhead{[days]} &
\colhead{[BKJD]} &
\colhead{[days]} &
\colhead{[BKJD]} &
\colhead{[hours]} &
\colhead{[minutes]} &
\colhead{}
}
\startdata
K2-52\,b     &   $3.534890^{+0.000187}_{-0.000181}$     &  $2063.02742^{+0.00190}_{-0.00184}$     &   $3.535055^{+0.000017}_{-0.000016}$     &  $2063.02618^{+0.00108}_{-0.00111}$     &      3.1     &        17     &       11x  \\
K2-53\,b     &  $12.207253^{+0.000940}_{-0.000971}$     &  $2063.38751^{+0.00400}_{-0.00373}$     &  $12.207720^{+0.000123}_{-0.000113}$     &  $2063.38582^{+0.00189}_{-0.00186}$     &      4.7     &        35     &        8x  \\
205084841.01     &  $11.310159^{+0.000564}_{-0.000557}$     &  $2060.85877^{+0.00224}_{-0.00221}$     &  $11.310099^{+0.000045}_{-0.000045}$     &  $2060.85913^{+0.00110}_{-0.00109}$     &      3.0     &        14     &       12x  \\
K2-289\,b     &  $13.157344^{+0.000622}_{-0.000634}$     &  $2064.14222^{+0.00205}_{-0.00198}$     &  $13.156969^{+0.000047}_{-0.000051}$     &  $2064.14325^{+0.00097}_{-0.00094}$     &      2.9     &        13     &       13x  \\
K2-174\,b     &  $19.564172^{+0.000996}_{-0.000966}$     &  $2250.77803^{+0.00100}_{-0.00099}$     &  $19.562307^{+0.000078}_{-0.000076}$     &  $2250.77917^{+0.00095}_{-0.00094}$     &      2.8     &        13     &       13x  \\
K2-87\,b     &   $9.726793^{+0.000663}_{-0.000646}$     &  $2239.30189^{+0.00229}_{-0.00242}$     &   $9.726618^{+0.000055}_{-0.000055}$     &  $2239.30232^{+0.00171}_{-0.00171}$     &      3.8     &        19     &       12x  \\
K2-90\,b     &  $13.733225^{+0.001206}_{-0.001187}$     &  $2245.66111^{+0.00185}_{-0.00190}$     &  $13.733314^{+0.000083}_{-0.000099}$     &  $2245.66112^{+0.00170}_{-0.00172}$     &      4.9     &        22     &       13x  \\
K2-124\,b     &   $6.413721^{+0.000271}_{-0.000272}$     &  $2309.18063^{+0.00187}_{-0.00182}$     &   $6.413651^{+0.000036}_{-0.000031}$     &  $2309.18106^{+0.00105}_{-0.00107}$     &      2.3     &        17     &        8x  \\

\enddata
\tablecomments{BKJD is the Barycentric Julian Date offset by the beginning of the \kepler mission, i.e. BJD--$2454833$. $^{\dagger}$Relative improvement in period measurement precision from our joint analysis of the \ktwo and \spitzer data, as compared to \ktwo-only.}
\end{deluxetable*}

\subsection{Stellar parameters}
\label{sec:analysis-stellar}

\begin{deluxetable*}{lcccccccc}
\tabletypesize{\scriptsize}
\tablecaption{\gaia DR2 parallax, 2MASS photometry, and spectroscopic priors used to estimate stellar parameters \label{tab:stellar_priors}}
\tablehead{EPIC	&	$\pi$	& $J$ & $H$ & $K$ &	\teff & \logg	&	\feh & Note \\
 & [mas] & [mag] & [mag] & [mag] & [K] & [cgs] & [dex] & }
\startdata
203776696 & 0.9394$\pm$0.1169 & 12.729$\pm$0.026 & 12.126$\pm$0.021 & 11.853$\pm$0.019 & $\hdots$ & $\hdots$ & $\hdots$ & $\hdots$ \\
204890128 & 7.2173$\pm$0.1161 & 10.306$\pm$0.024 & 9.826$\pm$0.024 & 9.664$\pm$0.021 & 5278$\pm$100 & 4.546$\pm$0.100 & -0.03$\pm$0.06 & 1 \\
205084841 & 1.1594$\pm$0.1117 & 13.443$\pm$0.026 & 12.797$\pm$0.026 & 12.612$\pm$0.027 & $\hdots$ & $\hdots$ & $\hdots$ & $\hdots$ \\
205686202 & 3.6170$\pm$0.1037 & 11.435$\pm$0.021 & 10.847$\pm$0.022 & 10.641$\pm$0.021 & 5365$\pm$110 & $\hdots$ & 0.16$\pm$0.08 & 3 \\
210558622 & 9.9783$\pm$0.1062 & 10.231$\pm$0.021 & 9.649$\pm$0.022 & 9.496$\pm$0.017 & 4310$\pm$70 & $\hdots$ & 0.11$\pm$0.09 & 2 \\
210731500 & 2.0071$\pm$0.1040 & 11.811$\pm$0.020 & 11.359$\pm$0.022 & 11.197$\pm$0.019 & 5694$\pm$100 & 4.067$\pm$0.100 & 0.36$\pm$0.06 & 1 \\
210968143 & 7.4234$\pm$0.1026 & 11.168$\pm$0.023 & 10.495$\pm$0.027 & 10.360$\pm$0.020 & 4465$\pm$100 & 4.553$\pm$0.100 & -0.25$\pm$0.06 & 1 \\
212154564 & 7.0953$\pm$0.1198 & 12.838$\pm$0.023 & 12.227$\pm$0.022 & 11.975$\pm$0.018 & 3443$\pm$70 & $\hdots$ & -0.13$\pm$0.09 & 2 \\

\enddata
\tablecomments{1: Keck/HIRES {\tt SpecMatch-syn}. 2: Keck/HIRES {\tt SpecMatch-emp}. 3: Subaru/HDS {\tt SpecMatch-emp}.}
\end{deluxetable*}

\begin{deluxetable*}{lccccccc}
\tabletypesize{\scriptsize}
\tablecaption{Stellar parameters \label{tab:stellar_posteriors}}
\tablehead{EPIC	&	\teff	& \logg & \feh & \mstar &	\rstar & distance	&	\rhostar \\
 & [K] & [cgs] & [dex] & [\msun] & [\rsun] & [pc]	&	[cgs] }
\startdata
203776696  &  $7147^{+483}_{-506}$  &  $3.979^{+0.087}_{-0.085}$  &   $0.055^{+0.154}_{-0.139}$  &  $1.691^{+0.157}_{-0.143}$  &  $2.192^{+0.280}_{-0.232}$  &  $1041.7^{+132.7}_{-104.1}$  &  $0.22^{+0.08}_{-0.06}$  \\
204890128  &    $5263^{+88}_{-86}$  &  $4.555^{+0.022}_{-0.028}$  &  $-0.037^{+0.052}_{-0.053}$  &  $0.851^{+0.032}_{-0.038}$  &  $0.807^{+0.016}_{-0.015}$  &       $139.1^{+2.2}_{-2.0}$  &  $2.29^{+0.15}_{-0.18}$  \\
205084841  &  $6245^{+369}_{-284}$  &  $4.271^{+0.062}_{-0.066}$  &   $0.019^{+0.150}_{-0.175}$  &  $1.168^{+0.120}_{-0.094}$  &  $1.313^{+0.126}_{-0.113}$  &     $860.4^{+84.1}_{-76.4}$  &  $0.73^{+0.18}_{-0.16}$  \\
205686202  &    $5529^{+77}_{-74}$  &  $4.393^{+0.024}_{-0.022}$  &   $0.162^{+0.087}_{-0.074}$  &  $0.953^{+0.037}_{-0.051}$  &  $1.025^{+0.030}_{-0.028}$  &       $270.7^{+7.6}_{-7.1}$  &  $1.24^{+0.10}_{-0.09}$  \\
210558622  &    $4455^{+31}_{-29}$  &  $4.625^{+0.017}_{-0.017}$  &   $0.113^{+0.065}_{-0.071}$  &  $0.700^{+0.025}_{-0.023}$  &  $0.676^{+0.008}_{-0.008}$  &       $100.1^{+1.0}_{-1.0}$  &  $3.21^{+0.15}_{-0.14}$  \\
210731500  &    $5747^{+58}_{-49}$  &  $4.208^{+0.032}_{-0.035}$  &   $0.328^{+0.051}_{-0.045}$  &  $1.164^{+0.042}_{-0.044}$  &  $1.401^{+0.083}_{-0.062}$  &     $503.6^{+29.9}_{-21.9}$  &  $0.59^{+0.07}_{-0.08}$  \\
210968143  &    $4484^{+58}_{-50}$  &  $4.653^{+0.017}_{-0.014}$  &  $-0.225^{+0.053}_{-0.053}$  &  $0.629^{+0.022}_{-0.018}$  &  $0.619^{+0.009}_{-0.010}$  &       $134.9^{+1.7}_{-1.7}$  &  $3.74^{+0.19}_{-0.14}$  \\
212154564  &    $3570^{+63}_{-70}$  &  $4.858^{+0.012}_{-0.088}$  &   $0.023^{+0.061}_{-0.106}$  &  $0.390^{+0.013}_{-0.042}$  &  $0.388^{+0.016}_{-0.009}$  &       $140.4^{+2.3}_{-2.3}$  &  $9.58^{+0.47}_{-2.12}$  \\

\enddata
\end{deluxetable*}

We used the Python package {\tt isochrones} to infer a uniform set of stellar parameters using priors from spectroscopy (when available), 2MASS $JHK$ photometry \citep{2006AJ....131.1163S}, and {\it Gaia} DR2 parallaxes \citep{2016A&A...595A...1G, 2018A&A...616A...1G}. We list the inputs to {\tt isochrones} in \autoref{tab:stellar_priors}. We sampled the posteriors using MultiNest \citep{2013arXiv1306.2144F} in conjunction with the MIST stellar evolution models \citep{2016ApJ...823..102C}. The resulting posteriors are listed in \autoref{tab:stellar_posteriors}.

\subsection{Planet validation}
\label{sec:analysis-validation}

The open-source Python package \vespa \citep{2015ascl.soft03011M} has been used to validate planets via statistical false positive probabilities (FPPs) in numerous recent works \citep[e.g.][]{2015ApJ...809...25M, 2016ApJ...822...86M}, and is similar to previous methods developed for \kepler and \corot, such as {\tt BLENDER} \citep{2011ApJ...727...24T} and {\tt PASTIS} \citep{2014MNRAS.441..983D}. Cr16 used \vespa to compute FPPs for planet candidates from \ktwo's first five observing campaigns, including those we analyze here (with the exception of K2-124\,b). \vespa uses the {\tt TRILEGAL} Galaxy model \citep{2005A&A...436..895G} to simulate populations of false positive (FP) scenarios (i.e. eclipsing binaries, background eclipsing binaries, and hierarchical triple systems) and then compares these to the observed phase-folded light curve to compute statistical likelihoods for each type of false FP, as well as the planetary scenario. For a more detailed description of how we use \vespa, see Cr16.

In common practice among planet hunters, the ``by-eye" transit shape is used as a first line of defense in the initial stages of vetting newly detected planet candidates in transit surveys: eclipsing binaries are usually more obviously ``V-shaped" than a planetary transit, as well as being deeper (in the absence of significant dilution). However, the 30 minute cadence of \ktwo can make the transits of real planets appear more V-shaped, so a more quantitative assessment based on precise parameter estimates is more reliable. The ratio of the full transit duration ($T_{23}$) to the total transit duration ($T_{14}$) is directly determined by the transit geometry, regardless of dilution from any additional sources within the photometric aperture \citep{2003ApJ...585.1038S}. We denote this ratio $\eta$, the transit ``shape" -- a number which scales between 0 and 1, where 0 corresponds to a ``V-shaped" transit and 1 corresponds to a ``box-shaped" transit:
\begin{align}
\label{eqn:shape}
\eta = \frac{T_{23}}{T_{14}}
\end{align}
The transit shape thus sets an upper limit on $R_p/R_{\star}$:
\begin{align}
\label{eqn:rpmax}
R_{p,max} / R_{\star} = \frac{1 - \eta}{1 + \eta}
\end{align}

As $\eta \rightarrow 1$, $R_{p}/R_{\star} \rightarrow R_{p,max}/R_{\star}$, meaning the constraint is stronger for more ``box-shaped" transits. This constraint can be particularly useful in the case of dilution from a known stellar companion (detected either photometrically or spectroscopically). \vespa implicitly uses the shape information content of the phase-folded transit light curve to compute the likelihoods of FP and planet models. Thus, an independent consideration of false positive scenarios is enabled by quantifying the shape $\eta$ of each transit. Furthermore, a constraint on the maximum allowed dilution results directly from comparison of the observed values of $R_p/R_{\star}$ and $R_{p,max}/R_{\star}$. The transit shape $\eta$ and the maximum radius ratio $R_{p,max}/R_{\star}$ are listed in \autoref{tab:params}, along with other parameters of interest.

Another useful constraint derived from the transit fit is the estimate of the mean stellar density. We compute this estimate of the stellar density directly from the observed transit light curve fit parameters using equation 4 of \citet{2014MNRAS.440.2164K}:
\begin{align}
\label{eqn:rho}
\rho_{\star,LC} = \frac{3\pi(a/R_{\star})^3}{GP^2}
\end{align}
This estimate can then be compared directly to independent estimates of the mean stellar density, i.e. from spectroscopy. Significant disagreement could arise from the violation of any of the assumptions inherent to this estimate (i.e. non-negligible blending, a non-circular orbit, $M_p \sim M_{\star}$).

\section{Discussion}
\label{sec:discussion}

\begin{figure}
\centering
\includegraphics[width=0.45\textwidth,trim={0.5cm 0cm 0.5cm 0cm}]{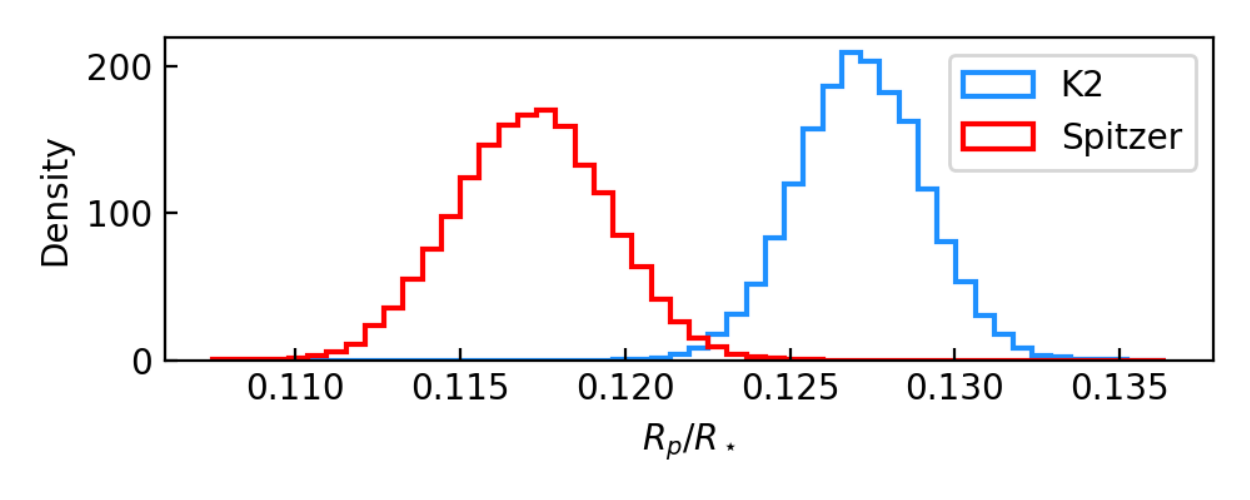}
\includegraphics[width=0.45\textwidth,trim={0.5cm 0cm 0.5cm 0cm}]{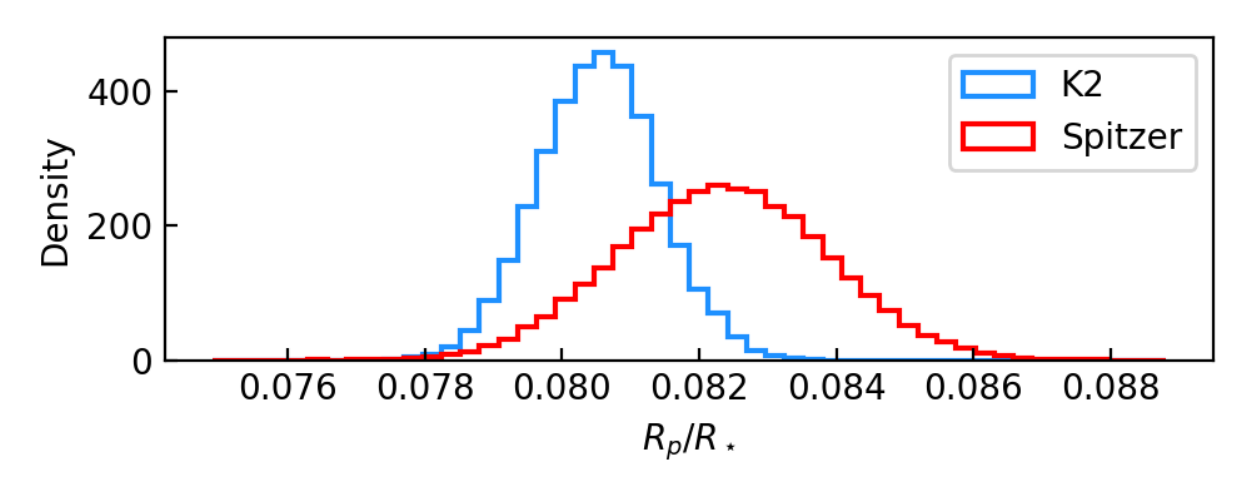}
\caption{Marginalized posterior distributions of $R_p/\rstar$ in the \kepler and \spitzer bandpasses for EPIC\,205084841.01 (top) and K2-289\,b (bottom). In the case of EPIC\,205084841.01, the disagreement casts doubt on the planetary hypothesis, while the agreement in the case of K2-289\,b is consistent with a planet.}
\label{fig:hist4841-6202}
\end{figure}

\subsection{Validation}
\label{sec:discussion-validation}

The planets in our sample with established ``validated'' dispositions (K2-52\,b, K2-53\,b, K2-87\,b, K2-90\,b, Cr16; K2-124\,b, \citealt{2017AJ....154..207D}, \citealt{2018arXiv181004074L}; K2-174\,b, \citealt{2018AJ....155..136M}) do not exhibit suspiciously chromatic transit depths that would indicate they are actually false positives. With the exception of K2-174, our transit analyses (without density priors) yield mean stellar densities ($\rho_{\star,LC}$) within 2$\sigma$ of the values from our independent assessment (\autoref{sec:analysis-stellar}). This agreement provides an additional layer of confidence in the disposition of these planets, as well as the quality of our transit analysis and stellar characterization.

On the other hand, K2-174\,b yields $\rho_{\star,LC} = 0.53^{+0.23}_{-0.21}$ \gcc, which is $\sim$10$\sigma$ discrepant with the value we derive for the host star, $\rho_\star = 3.21^{+0.15}_{-0.14}$ \gcc. Furthermore, the addition of a density prior does not significantly change this result, which indicates that the data strongly constrain the density to this value. We interpret this result as an indication of eccentricity in the system, which could potentially be measured via RVs (see \autoref{sec:discussion-potential}).

In the next two subsections we examine the dispositions of the two previously un-validated planet candidates in our sample, EPIC\,205084841.01 and K2-289\,b.

\subsubsection{EPIC\,205084841.01}

EPIC\,205084841.01 does not warrant validation at present, but neither should it be considered a ``confirmed'' false positive; the current disposition of planet candidate is appropriate. Using the contrast curve derived from our Keck/NIRC2 AO image (see \autoref{fig:ao}), along with the {\tt isochrones} inputs listed in \autoref{tab:stellar_priors}, we compute an FPP of 15.4\% for this target with \vespa, somewhat higher than the FPP of 2.7\% reported by Cr16, and significantly above the commonly used validation threshold of 1\%. We discuss below several considerations pertinent to the disposition of this system.

Caution is warranted by the candidate's large radius of $\sim$17 \rearth, which is comparable to the radii of low mass stars \citep{2017ApJ...847L..18S}. Additionally, there is moderate tension between $R_p/\rstar$ in the \kepler and \spitzer bandpasses (see \autoref{fig:hist4841-6202}). The posterior of $R_p/\rstar$ is $\sim$3.3$\sigma$ larger in the \kepler bandpass, which is consistent with an occultation by a lower mass star. Although the FPP implies the candidate is more likely a planet than a false positive, \vespa does not take into account chromaticity of transit depth. The second-most likely scenario reported by \vespa is an EB, and \vespa essentially rules out a hierarchical or background EB scenario.

Assuming the candidate is actually an EB with the same orbital period, a secondary eclipse could be measurable in the \ktwo data; the absence of any apparent secondary eclipses (deeper than $\sim$0.1\% at any phase, see \autoref{fig:k2-4841}) is therefore suggestive of an eccentric orbit. Our transit analysis (without a density prior) yields $\rho_{\star,LC} = 1.43^{+0.29}_{-0.34}$ \gcc, which is 1.8$\sigma$ higher than the value we measure for the host star (see \autoref{tab:stellar_posteriors}) and thus modestly suggestive of eccentricity. There is no apparent variation in the transit depth and the transit geometry implies a radius ratio much smaller than unity ($\lesssim$20\%, see \autoref{tab:params}), so an EB at twice the estimated orbital period is unlikely. We conclude that EPIC\,205084841.01 is potentially a false positive caused by the eclipses of a low mass star in an eccentric orbit, but further observations are required to confirm this scenario (e.g. RV monitoring); based on its FPP alone, the candidate is more likely to be an inflated gas giant planet.

\subsubsection{K2-289\,b}

Cr16 reported a FPP of $1.3 \times 10^{-11}$ for this candidate, but did not validate the candidate due to the presence of a nearby stellar companion revealed by AO imaging. Using the contrast curve for this star in \autoref{fig:ao} and the {\tt isochrones} inputs listed in \autoref{tab:stellar_priors}, we find a higher (but still rather low) FPP of $3.7 \times 10^{-4}$. The companion is at a separation of 0.8\arcsec, and is 3.8 magnitudes fainter (in $K$ band) than the primary star. The 1.2\arcsec\, pixel scale of IRAC is thus insufficient to resolve the companion in our follow-up \spitzer transit photometry. Intriguingly, the companion was not detected by {\it Gaia} DR2 but K2-289 is listed with zero excess astrometric noise, which is often associated with binarity (see e.g. \citealt{2018RNAAS...2b..20E}). This underscores the need for high resolution imaging, even in the {\it Gaia} era.

From our AO imaging (\autoref{sec:observations-imaging}), we compute deblended $J$ band magnitudes of $11.461 \pm 0.024$ and $15.554 \pm 0.055$ for the primary and secondary stars, respectively; in $K$ band the deblended magnitudes are $10.674 \pm 0.021$ and $14.446 \pm 0.024$. The $J-K$ color of the secondary star is thus $1.108 \pm 0.060$, which is consistent with a late M dwarf.

From the observed transit depth, an eclipsing binary (EB) with 100\% eclipse depth that is less than 5.4 magnitudes fainter than the primary could reproduce the observed transit depth. However, the $J-K$ color of the companion suggests it is a late M dwarf, so the contrast in the \kepler bandpass should be $\gtrsim$4 magnitudes greater than in $K$ band. In this likely scenario, we can rule out the secondary as the source of the signal, as the dilution from the primary star in the \kepler bandpass would require a depth larger than 100\%. Furthermore, we can leverage our measurement of the transit geometry to eliminate considerations of the secondary star's spectral type. From our joint analysis of the \ktwo and \spitzer data, we have the constraint $R_{p,max}/R_{\star} < 0.153$ (5$\sigma$), which corresponds to a maximum undiluted transit depth of $\sim$2.3\%. Given the observed transit depth, this translates to an upper limit on the amount of dilution of about one magnitude. The secondary star would thus need to have a non-physical color of $V-K \lesssim -2.8$ to be the source of the signal. We therefore conclude that the observed signal comes from the primary star.

From the above considerations we can confidently rule out scenarios in which the signal comes from the secondary source, i.e. the signal is due to a BEB or eclipsing hierarchical triple (HEB). We now consider the possibility that the primary is itself an eclipsing binary. While low mass stars can potentially be roughly Jovian in size, such massive bodies would induce a large amplitude Doppler signal in time series RV measurements of the primary star, as well as potentially significant secondary eclipses.

The RVs obtained with Subaru/HDS (see \autoref{sec:observations-spectroscopy}) show no significant variation on a timescale of $\sim$3 days, ruling out the possibility that K2-289 is an eclipsing binary. Assuming a circular orbit, we used \radvel to estimate the RV semi-amplitude exerted by the transiting companion as $K = 0.03^{+5.47}_{-0.03}$ m\,s$^{-1}$, which is consistent with a null detection. The semi-amplitude best-fit value and 3$\sigma$ upper limit are 11.8 m\,s$^{-1}$ and 46.1 m\,s$^{-1}$, respectively. This result is consistent with the \vespa result as well as our assessment of the \ktwo and \spitzer transit data, which both suggest a planetary origin for the observed transit signal, so we conclude this is a valid planet.

Since we do not account for the dilution from the companion in our transit fits, the planet radius we measure may be underestimated. However, the delta-magnitude of 3.82 (in $K$ band), implies a planet radius only $\sim$1.5\% larger than we list in \autoref{tab:planets}, which is a factor of $\sim$6 smaller than the precision of our planet radius measurement. Furthermore, the dilution is likely to be lower in \kepler band than $K$ band, given the secondary is probably a bound late-type companion. Future studies may yield precise enough measurements of $R_p/R_{\star}$ that dilution from the companion will need to be accounted for.

\subsection{Potential for characterization}
\label{sec:discussion-potential}

\begin{deluxetable*}{lcccccccc}
\tabletypesize{\scriptsize}
\tablecaption{Derived planet properties \label{tab:planets}}
\tablehead{Name & $R_p$ & $M_p$ & $K_\mathrm{pred}$ & $a$ & \Teq & $g$ & H & $\delta_{\mathrm{TS}}$ \\
  & [\rearth] & [\mearth] & [m\,s$^{-1}$] & [AU] & [K] & [$g_\oplus$] & [km] & [ppm]}
\startdata
K2-52\,b     &  $18.0 \pm 2.2$     &  $128.6 \pm 109.9$     &  $38.1 \pm 32.6$     &  $0.054 \pm 0.002$     &  $2004 \pm 184$     &  $0.4$     &  $2129$     &    $1049$  \\
K2-53\,b     &   $2.6 \pm 0.1$     &      $9.2 \pm 2.5$     &    $2.8 \pm 0.8$     &  $0.098 \pm 0.001$     &    $665 \pm 13$     &  $1.3$     &   $211$     &     $112$  \\
205084841.01     &  $17.5 \pm 1.7$     &   $123.4 \pm 96.3$     &  $31.8 \pm 25.0$     &  $0.104 \pm 0.003$     &    $979 \pm 73$     &  $0.4$     &  $1029$     &    $1374$  \\
K2-289\,b     &   $9.1 \pm 0.3$     &    $48.6 \pm 18.6$     &   $13.6 \pm 5.2$     &  $0.107 \pm 0.001$     &    $753 \pm 16$     &  $0.6$     &   $545$     &     $623$  \\
K2-174\,b     &   $2.6 \pm 0.1$     &      $9.2 \pm 2.5$     &    $2.8 \pm 0.8$     &  $0.126 \pm 0.001$     &     $455 \pm 5$     &  $1.3$     &   $145$     &     $110$  \\
K2-87\,b     &   $7.6 \pm 0.4$     &    $37.9 \pm 12.7$     &   $10.3 \pm 3.4$     &  $0.094 \pm 0.001$     &    $979 \pm 28$     &  $0.7$     &   $633$     &     $323$  \\
K2-90\,b     &   $2.6 \pm 0.1$     &      $9.0 \pm 2.5$     &    $3.3 \pm 0.9$     &  $0.096 \pm 0.001$     &     $502 \pm 8$     &  $1.4$     &   $157$     &     $139$  \\
K2-124\,b     &   $2.9 \pm 0.1$     &     $10.5 \pm 2.6$     &    $6.8 \pm 1.8$     &  $0.049 \pm 0.002$     &    $442 \pm 17$     &  $1.2$     &   $151$     &     $383$  \\

\enddata
\tablecomments{$M_p$ comes from the probabilistic mass-radius relation of \citet{2016ApJ...825...19W}; we apply this relation uniformly for the sake of homogeneity, but note that $M_p$ (and its uncertainty) may be underestimated for the larger planets in our sample. $K_\mathrm{pred}$ is the predicted RV semi-amplitude, $a$ is semi-major axis, and \Teq is equilibrium temperature assuming a Bond albedo of 0.3; the uncertainties in these parameters propagate from the formal transit and stellar parameter estimates but are likely underestimated due to uncertainties in the underlying models and assumptions. $g$ is surface gravity, H is atmospheric scale height, and $\delta_{\mathrm{TS}}$ is the predicted amplitude of atmospheric features accessible via transmission spectroscopy; these estimates have large uncertainties owing to a number of factors, such as uncertainties in the stellar parameters, unknown planet masses, and assumed atmospheric compositions.}
\end{deluxetable*}

In order to assess the potential for future characterization studies of these planets, we computed their physical properties using the parameter estimates from our transit analyses in \autoref{tab:params} and \autoref{tab:ephemeris}, as well as the stellar parameters in \autoref{tab:stellar_posteriors}. We first computed updated planet radii and used the probabilistic mass-radius relation of \citet{2016ApJ...825...19W} to predict the masses of these planets. We then computed the orbital semi-major axis (in physical units), predicted RV semi-amplitude (assuming circular orbits), equilibrium temperature (assuming a Bond albedo of 0.3), surface gravity, and atmospheric scale height (assuming a Hydrogen-dominated atmosphere, i.e. a mean molecular weight of 2). Finally, we used the formalism of \citet{2009ApJ...690.1056M} to estimate the amplitude of features in the planets' transmission spectra. We list these derived planet properties in \autoref{tab:planets}.

K2-53 and K2-174 are moderately bright in the optical ($Kp \sim 12$) and each host a $R_p\sim2.7$\rearth planet with a predicted RV semi-amplitudes of $K_\mathrm{pred} \sim 3$ m\,s$^{-1}$. These are thus potentially good targets for precision RV mass measurements with current optical spectrographs such as HIRES and HARPs, which would enable studies of the densities and bulk composition of temperate sub-Neptunes.
K2-289\,b and K2-87\,b are both sub-Saturns orbiting moderately IR-bright stars ($J < 12$) with $K_\mathrm{pred} = 13.6 \pm 5.2$ and $K_\mathrm{pred} = 10.3 \pm 3.4$ m\,s$^{-1}$, respectively, making them good targets for one of the high precision ``red'' (IR) spectrographs currently under development (e.g. Subaru/IRD \citep{2012SPIE.8446E..1TT}, Spirou \citep{2014SPIE.9147E..15A}, and HPF \citep{2015AAS...22525823M}). Mass measurements for these planets would add crucial data points to the relatively small population of similarly sized planets with well-measured densities, which exhibit a diverse range of core mass fractions \citep{2017AJ....153..142P}. K2-90 and K2-124 are similarly IR-bright, but host temperate sub-Neptunes with $K_\mathrm{pred} = 3-6$ m\,s$^{-1}$, making them also potentially interesting targets for red spectrographs.

Due to the small size and moderate infrared brightness of the host star, K2-124\,b could be an interesting target for atmospheric studies via transmission spectroscopy with \jwst. The predicted amplitude of features in the planet's transmission spectrum would be roughly 350 ppm (with large uncertainties owing to the unknown planet mass and atmospheric mean molecular weight). By combining multiple transit observations, these features should be detectable with \jwst. This 2.5\rearth planet receives only $\sim$8 times the incident radiation as the Earth, so such studies would probe the atmospheric properties of temperate sub-Neptunes.

The combination of \ktwo and \spitzer data in this work demonstrates the synergy between transit detection and follow-up with space-based instruments. Because \tess has a pixel scale five times larger than \kepler (21\arcsec\ vs. 4\arcsec), the frequency of stellar blends will likely be significantly larger, which will increase the utility of transit follow-up. The upcoming \cheops mission has a similar pixel scale to \spitzer, and will thus prove useful for \tess follow-up in much the same way that \spitzer has for \ktwo.

\section{Summary}
\label{sec:summary}

We have used \spitzer to observe the transits of eight planet candidates discovered by \ktwo, and we perform a global analysis of the light curves from both telescopes. The value of \spitzer follow-up transit observations of \ktwo candidates is two-fold -- the high cadence infrared light curves allow a finer sampling of the transit shape which is less confounded by uncertain limb darkening, and the measurement of an additional transit time at a later epoch yields a more precise estimate of the ephemeris. Our follow-up transit observations with \spitzer demonstrate the utility of transit follow-up observations for planet validation and ephemeris refinement, paving the way for future RV and atmospheric studies. These observations reduce mid-2021 transit timing uncertainties by $\sim$90\%, thus enabling efficient scheduling with \jwst.

\acknowledgements
This work is based in part on observations made with the \spitzer Space Telescope, which is operated by the Jet Propulsion Laboratory, California Institute of Technology (Caltech) under contract with the National Aeronautics and Space Administration (NASA). This paper includes data collected by the \ktwo mission. Funding for the \ktwo mission is provided by the NASA Science Mission directorate. This work benefited from the Exoplanet Summer Program in the Other Worlds Laboratory (OWL) at the University of California, Santa Cruz, a program funded by the Heising-Simons Foundation. E. S. is supported by a postgraduate scholarship from the Natural Sciences and Engineering Research Council of Canada. E. A. P. acknowledges support by NASA through a Hubble Fellowship grant awarded by the Space Telescope Science Institute, which is operated by the Association of Universities for Research in Astronomy, Inc., for NASA, under contract NAS 5-26555. A. W. H. acknowledges support for our \ktwo team through a NASA Astrophysics Data Analysis Program grant. A. W. H. and I. J. M. C. acknowledge support from the K2 Guest Observer Program. This work was performed [in part] under contract with the Jet Propulsion Laboratory (JPL) funded by NASA through the Sagan Fellowship Program executed by the NASA Exoplanet Science Institute. This research has made use of the NASA Exoplanet Archive, which is operated by Caltech, under contract with NASA's Exoplanet Exploration Program. This work made use of the SIMBAD database (operated at CDS, Strasbourg, France) and NASA’s Astrophysics Data System Bibliographic Services. This research made use of the Infrared Science Archive, which is operated by the Caltech, under contract with NASA. Portions of this work were performed at the California Institute of Technology under contract with NASA. Some of the data presented herein were obtained at the W.M. Keck Observatory, which is operated as a scientific partnership between Caltech, the University of California, and NASA. The authors wish to extend special thanks to those of Hawai'ian ancestry, on whose sacred mountain of Maunakea we are privileged to be quests. We are most fortunate to have the opportunity to conduct observations from this mountain.

\facilities{\kepler, \spitzer, {\it Gaia}, Keck (NIRC2, HIRES), Subaru (HDS)}

\software{{\tt numpy} \citep{numpy}, {\tt scipy} \citep{scipy}, {\tt matplotlib} \citep{Hunter:2007}, {\tt lmfit} \citep{newville_2014_11813}, {\tt emcee} \citep{emcee}, \celerite \citep{2017AJ....154..220F}, {\tt batman} \citep{2015PASP..127.1161K}, {\tt isochrones} \citep{2015ascl.soft03010M}, {\tt vespa}, \citep{2015ascl.soft03011M}, \ktwophot \citep{2015ApJ...811..102P}, \specmatchsyn \citep{Petigura15}, \specmatchemp \citep{2017ApJ...836...77Y}, \radvel \citep{2018PASP..130d4504F}, {\tt IRAF} \citep{1986SPIE..627..733T,1993ASPC...52..173T}}

\bibliography{ref.bib}

\appendix

\section{Limb darkening}
\label{sec:ld}

\begin{deluxetable*}{l|cc|cc}
\tablecolumns{5}
\tabletypesize{\scriptsize}
\tablecaption{Limb darkening priors\label{tab:ld}}
\tablehead{
\colhead{} &
\multicolumn{2}{c}{\ktwo} &
\multicolumn{2}{c}{\spitzer} \\
\colhead{Name} &
\colhead{$q_{1}$} &
\colhead{$q_{2}$} &
\colhead{$q_{1}$} &
\colhead{$q_{2}$}
}
\startdata
K2-52 & $\mathcal{N}(0.339,0.024)$ & $\mathcal{N}(0.237,0.030)$ & $\mathcal{N}(0.026,0.005)$ & $\mathcal{N}(0.174,0.022)$ \\
K2-53 & $\mathcal{N}(0.490,0.012)$ & $\mathcal{N}(0.372,0.016)$ & $\mathcal{N}(0.047,0.002)$ & $\mathcal{N}(0.198,0.019)$ \\
205084841 & $\mathcal{N}(0.392,0.034)$ & $\mathcal{N}(0.271,0.028)$ & $\mathcal{N}(0.035,0.005)$ & $\mathcal{N}(0.165,0.023)$ \\
K2-289 & $\mathcal{N}(0.481,0.011)$ & $\mathcal{N}(0.339,0.012)$ & $\mathcal{N}(0.043,0.002)$ & $\mathcal{N}(0.203,0.023)$ \\
K2-174 & $\mathcal{N}(0.561,0.007)$ & $\mathcal{N}(0.463,0.005)$ & $\mathcal{N}(0.055,0.001)$ & $\mathcal{N}(0.213,0.019)$ \\
K2-87 & $\mathcal{N}(0.453,0.008)$ & $\mathcal{N}(0.327,0.007)$ & $\mathcal{N}(0.043,0.001)$ & $\mathcal{N}(0.153,0.015)$ \\
K2-90 & $\mathcal{N}(0.553,0.003)$ & $\mathcal{N}(0.448,0.003)$ & $\mathcal{N}(0.058,0.001)$ & $\mathcal{N}(0.205,0.021)$ \\
K2-124 & $\mathcal{N}(0.530,0.023)$ & $\mathcal{N}(0.266,0.029)$ & $\mathcal{N}(0.038,0.001)$ & $\mathcal{N}(0.082,0.036)$ \\
\enddata
\end{deluxetable*}

We used Gaussian limb darkening priors in our transit analysis of the \ktwo and \spitzer data (see \autoref{sec:analysis-transitmodel}). We use our stellar parameter estimates to determine these priors, so the more uncertain our knowledge of the host star, the less informative these priors are. Our Gaussian priors have typical widths of $\sim$10\%, which is comparable to the uncertainty from the stellar limb darkening models \citep[e.g.][]{2013A&A...549A...9C, 2013A&A...560A.112M}. In practice, these priors do not have a large effect on the final results, especially in the case of the \spitzer data, where limb darkening is almost negligible at 4.5 \microns. Conversely, the light curves are not sufficient to constrain limb darkening empirically -- the photometric precision of the \spitzer data is too low, and the cadence of the \ktwo data is too long. As a result, the posteriors of the limb darkening parameters are nearly identical to the priors. We list the limb darkening priors used in our analyses in \autoref{tab:ld}.

\begin{deluxetable}{lcc}
\tabletypesize{\scriptsize}
\tablecolumns{3}
\tablecaption{The effect of the choice of limb darkening law on estimates of $R_p$/\rstar, for a range of host star effective temperatures. The subscripts $K$ and $S$ denote the \kepler and \spitzer (4.5\microns) bandpasses, respectively. \label{tab:ldlaw}}
\tablehead{
\colhead{Name} &
\colhead{$R_{p,K}$} &
\colhead{$R_{p,S}$} \\
\colhead{} &
\colhead{[\rstar]} &
\colhead{[\rstar]}
}
\startdata
\multicolumn{3}{l}{{\it linear}} \\
205084841.01 & $0.12171^{+0.00231}_{-0.00187}$ & $ 0.11602^{+0.00237}_{-0.00225}$  \\
K2-174\,b & $0.03605^{+0.00244}_{-0.00179}$ & $ 0.03892^{+0.00129}_{-0.00143}$  \\
K2-124\,b & $0.06791^{+0.00398}_{-0.00255}$ & $ 0.07005^{+0.00373}_{-0.00318}$  \\
\cline{1-3}
\multicolumn{3}{l}{{\it quadratic}} \\
205084841.01 & $0.12134^{+0.00208}_{-0.00179}$ & $ 0.11616^{+0.00223}_{-0.00248}$  \\
K2-174\,b & $0.03593^{+0.00301}_{-0.00160}$ & $ 0.03864^{+0.00129}_{-0.00133}$  \\
K2-124\,b & $0.06761^{+0.00333}_{-0.00226}$ & $ 0.07011^{+0.00317}_{-0.00342}$  \\
\cline{1-3}
\multicolumn{3}{l}{{\it square-root}} \\
205084841.01 & $0.12187^{+0.00215}_{-0.00202}$ & $ 0.11586^{+0.00246}_{-0.00232}$  \\
K2-174\,b & $0.03615^{+0.00284}_{-0.00183}$ & $ 0.03868^{+0.00131}_{-0.00130}$  \\
K2-124\,b & $0.06770^{+0.00367}_{-0.00243}$ & $ 0.06994^{+0.00326}_{-0.00329}$  \\
\cline{1-3}
\multicolumn{3}{l}{{\it logarithmic}} \\
205084841.01 & $0.12134^{+0.00197}_{-0.00172}$ & $ 0.11584^{+0.00246}_{-0.00229}$  \\
K2-174\,b & $0.03597^{+0.00260}_{-0.00180}$ & $ 0.03881^{+0.00124}_{-0.00137}$  \\
K2-124\,b & $0.06774^{+0.00374}_{-0.00219}$ & $ 0.07043^{+0.00324}_{-0.00340}$  \\
\enddata
\end{deluxetable}

The choice of limb darkening law can potentially affect the posteriors of system parameters, due to deficiencies in the parametrization of the underlying flux variations across the stellar disk \citep[e.g.][]{2016MNRAS.457.3573E, 2017AJ....154..111M}. We tested the effect of using a quadratic limb darkening law by repeating our analysis of the \ktwo and \spitzer data for three host stars spanning a range of effective temperatures, using a range of limb darkening laws. Specifically, we tested a total of four different limb darkening laws: linear, quadratic, square-root, and logarithmic. For each limb darkening law, we computed priors for the limb darkening coefficients via Monte Carlo interpolation of the appropriate tables of \citet{2012yCat..35460014C}, enabling uncertainties in stellar parameters to propagate to uncertainties in the limb darkening parameters for each law, as well as helping to ensure that only physical limb darkening solutions are considered during MCMC. We list the posteriors of $R_p$/\rstar in the \kepler and \spitzer 4.5\micron\ bands for EPIC 205084841 ($\sim$6250 K), K2-174 ($\sim$4450 K), and K2-124 ($\sim$3570 K) in \autoref{tab:ldlaw}. As no significant differences in the estimates of $R_p$/\rstar can be seen in either bandpass, we conclude that the choice of limb darkening law is not important in the case of these datasets. A combination of high photometric precision and high cadence, e.g. \kepler short cadence data, is likely necessary in order to observe significant dependence of transit parameter estimates on the choice of limb darkening law.

\end{document}